\documentclass[prd,aps,floats,floatfix,nofootinbib,preprintnumbers]{revtex4-1}
\usepackage{amsmath}
\usepackage{amsfonts}
\usepackage{graphicx}
\usepackage{slashed}
\usepackage{dcolumn}
\usepackage{bm}
\usepackage{amssymb}
\usepackage{latexsym}
\usepackage{color}
\usepackage{pdfpages}

\newcommand{\brac}[1]{\left( #1 \right)}
\newcommand{\pbrac}[1]{\left( #1 \right)}
\newcommand{\tbrac}[1]{\left[ #1 \right]}
\newcommand{\cbrac}[1]{\left\{ #1 \right\}}

\begin{document}

\title{Constraints on the Scalar Sector of \\ the Renormalizable Coloron Model}

\author{R. Sekhar Chivukula}
\email[]{sekhar@msu.edu}
\affiliation{Department of Physics and Astronomy, Michigan State University, East Lansing, Michigan 48824, USA}
\author{Arsham Farzinnia}
\email[]{farzinnia@tsinghua.edu.cn}
\affiliation{Institute of Modern Physics, Center for High Energy Physics Research, Tsinghua University, Beijing 100084, China}
\author{Jing Ren}
\email[]{renj08@mails.tsinghua.edu.cn}
\affiliation{Institute of Modern Physics, Center for High Energy Physics Research, Tsinghua University, Beijing 100084, China}
\author{Elizabeth H. Simmons}
\email[]{esimmons@msu.edu}
\affiliation{Department of Physics and Astronomy, Michigan State University, East Lansing, Michigan 48824, USA}

\date{\today}

\begin{abstract}
The renormalizable coloron model  is the minimal extension of the standard model color sector, in which the color gauge group is enlarged to $SU(3)_{1c} \times SU(3)_{2c}$. In this paper we discuss the constraints on this model derived from the requirements of vacuum stability, tree-level unitarity, electroweak precision measurements, from LHC measurements of the properties of the observed Higgs-like scalar boson, and from LHC limits on additional Higgs-like bosons decaying to dibosons. The combination of these theoretical and experimental considerations strongly constrains the allowed
parameter space.
 \end{abstract}

\maketitle

\section{Introduction}\label{intro}

The ATLAS \cite{Aad:2012tfa} and CMS \cite{Chatrchyan:2012ufa} experiments at the LHC have provided conclusive evidence for a scalar boson with Higgs-like propeties and a mass of approximately 125 GeV. While this discovery provides the first glimpse of physics at the TeV energy scale, many important questions (such as the origin of the Higgs vacuum expectation value and of the multitude of fermion masses and mixings) remain unanswered. Theories that address these questions require dynamics beyond that in the standard model.

One attractive possible feature of theories beyond the standard model is an extension to the color sector of the standard model. The simplest possiblity for this extension is to enlarge the QCD gauge group to $SU(3)_{1c} \times SU(3)_{2c}$, with ordinary color identified with the diagonal subgroup of this larger symmetry. 
Models in this category include topcolor
\cite{Hill:1991at}, the flavor-universal coloron \cite{Chivukula:1996yr}, chiral color \cite{Frampton:1987dn}, chiral color with unequal gauge couplings \cite{Martynov:2009en} and a newer flavor non-universal chiral color model \cite{Frampton:2009rk}. In addition to a set of  massive color octet vector bosons arising from the expanded color interactions (states we refer to here generically as colorons), there will necessarily be additional scalar states associated with breaking the extended color sector to ordinary QCD. These new colored states can have a large effect on the properties of the standard model Higgs boson \cite{Manohar:2006ga,Bai:2011aa,Dobrescu:2011aa,Kumar:2012ww,Chang:2012ta,Kribs:2012kz,Dorsner:2012pp,Huang:2012wn,Cao:2013wqa}.

The renormalizable coloron model \cite{Hill:1993hs,Dicus:1994sw,Chivukula:1996yr,Bai:2010dj} is the minimal extension of the standard model incorporating an enlarged color gauge group that provides a framework for examining the interplay between the breaking of the electroweak and extended color gauge symmetries. In this paper we discuss the constraints on this model derived from the requirements of vacuum stability, tree-level unitarity, electroweak precision measurements, and from LHC measurements of the properties of the observed Higgs-like scalar boson. 

Our analysis demonstrates the interplay between these various constraints. Tree-level unitarity arguments constrain the masses of the scalar particles associated with breaking the extended color sector in relation to that sector's symmetry breaking scale.  The bounds derived from precision electroweak measurements strongly limit the amount by which the observed 125 GeV scalar mass eigenstate can mix with a gauge-singlet state from the color symmetry breaking sector. The observed production and decay properties of the 125 GeV scalar also constrain this mixing in a complementary regime, though in a manner that depends on the details of the model. These constraints on mixing, in turn, have consequences for the spectrum of scalar states. A summary of the range of parameters allowed in the renormalizable coloron model is displayed in Fig.~\ref{combo}.

Our work builds on many recent theoretical investigations. Closely related constraints can be obtained in models that contain a scalar singlet in addition to the standard model Higgs boson, see \cite{Pruna:2013bma} and references therein. Considerations of the mixing between the standard model Higgs boson and scalars from other sectors is an example of the ``Higgs Portal" introduced in \cite{Schabinger:2005ei,Barbieri:2005ri,Patt:2006fw}. Our discussion here is complementary to direct searches for the vectors \cite{Simmons:1996fz,Bertram:1998wf,ATLAS:2012pu,ATLAS:2012qjz,Chatrchyan:2013qha,CMS:kxa} or scalars \cite{Aaltonen:2013hya} present in the model, or to the theoretical investigations of this class of models based on their flavor couplings \cite{Manohar:2006ga,Gresham:2007ri,Arnold:2011ra,Gerbush:2007fe,Carpenter:2011yj,Chivukula:2013kw,Chivukula:2013hga}.

In the next section we review the renormalizable coloron model \cite{Hill:1993hs,Dicus:1994sw,Chivukula:1996yr,Bai:2010dj} and set our notation. The following section demonstrates the theoretical bounds from requiring vacuum stability and tree-level unitarity. The fourth section establishes the constraints on the scalar sector of the renormalizable coloron model from experimental results on precisely measured electroweak quantities, from the observed properties of the 125 GeV scalar boson, and from LHC limits on the additional Higgs-like bosons decaying to dibosons. The last section combines the individual analyses and includes a summary of our results.

\section{The Renormalizable Coloron Model}\label{model}

The renormalizable coloron model 
\cite{Hill:1993hs,Dicus:1994sw,Chivukula:1996yr,Bai:2010dj}
consists of a minimal renormalizable extension of the standard model (SM) color sector, in which the color gauge group is enlarged to $SU(3)_{1c} \times SU(3)_{2c}$. In this model, the spontaneous symmetry breaking of the SM electroweak sector is accompanied by a spontaneous breaking of the enlarged color gauge group to the diagonal subgroup $SU(3)_{c}$, which is identified with ordinary QCD.

Hence, the model contains, in addition to the usual massless gluon color-octet, a set of massive color-octet vector bosons, generically called \textit{colorons}. To facilitate the spontaneous breaking of the symmetry in the enhanced color sector, the theory includes a scalar
$(3,\bar{3})$ under the $SU(3)_{1c} \times SU(3)_{2c}$ interactions.
Under $SU(3)_c$  this boson includes a
gauge-singlet scalar, a gauge-singlet pseudo-scalar, a set of electroweak-singlet color-octet scalars, and a set of electroweak-singlet color-octet pseudoscalars which are ``eaten" by the massive colorons. The symmetry breaking in the extended color sector is induced by the CP-even singlet scalar component's developing a non-zero vacuum expectation value (VEV). In principle, this degree of freedom is capable of mixing with the SM electroweak Higgs boson, giving rise to potentially interesting phenomenology \cite{Kumar:2012ww}. Moreover, as described below, the renormalizable coloron model includes heavy spectator quarks \cite{Frampton:1987dn,Frampton:1987ut,Cvetic:2012kv,Chivukula:2013xla} which serve to cancel potential anomalies introduced by the chiral couplings of the quarks to the extended color gauge group.

\subsection{The Bosonic Sector}

The model is described by the Lagrangian
\begin{equation}\label{Lagr}
\mathcal{L} = \mathcal{L}_{\text{electroweak}} + \mathcal{L}_{\text{color}} + \mathcal{L}_{\text{scalar}} + {\cal L}_{\rm fermion} \ ,
\end{equation}
where
\begin{align}
\mathcal{L}_{\text{color}} =&\, - \frac{1}{2} \text{Tr} \left [ G_{1 \mu \nu} G_{1}^{ \mu \nu} \right ] - \frac{1}{2} \text{Tr} \left [ G_{2 \mu \nu} G_{2}^{ \mu \nu} \right ] + {\cal L}_{\rm gauge-fixing} + {\cal L}_{\rm ghost} \ , \label{Lcol} \\
\mathcal{L}_{\text{scalar}} =&\, D^\mu \phi^\dagger D_\mu \phi+{\rm Tr} \tbrac{D^\mu\Phi^{\dagger} D_\mu\Phi} - V(\phi,\Phi) \ . \label{Lscal}
\end{align}
In \eqref{Lcol}, $G_{1 \mu \nu}$ and $G_{2 \mu \nu}$ represent the field-strength tensors of the original $SU(3)_{1c}$ and $SU(3)_{2c}$ gauge bosons, respectively, with the corresponding couplings $g_{s_{1}}$ and $g_{s_{2}}$. The SM electroweak gauge sector is unaltered, and the field~$\phi$, in \eqref{Lscal}, is the SM Higgs doublet, responsible for electroweak symmetry breaking, which may be written in component form as
\begin{equation}\label{phi}
\phi= \frac{1}{\sqrt{2}}
\begin{pmatrix} i\sqrt{2}\,\pi^+ \\ v_h+h_0+i\pi^0 \end{pmatrix} \ .
\end{equation}
In this expression, $h_0$ is the SM Higgs boson with the associated VEV $v_h=246$~GeV, and $\pi^{0,\pm}$ are the usual electroweak Nambu-Goldstone bosons. The electroweak covariant derivative is defined in the standard way
\begin{equation}\label{EWcoder}
D_\mu \phi= \partial_\mu \phi - ig\, W_\mu^b \tau^b \phi + \frac{i}{2}g^\prime B_\mu \phi \qquad \pbrac{\tau^b \equiv \sigma^b/2} \ ,
\end{equation}
with $\sigma^{b}$ the Pauli matrices.

The $\Phi$ field in \eqref{Lscal}, on the other hand, is responsible for the spontaneous symmetry breaking in the enhanced color sector, and has the component form \cite{Bai:2010dj}
\begin{equation}\label{Phi}
\Phi = \frac{1}{\sqrt{6}} \pbrac{v_{s} + s_{0} + i {\cal A}} {\cal I}_{3\times 3} + \pbrac{G^a_H + i G^a_G}t^a \qquad \pbrac{t^a \equiv \lambda^a/2} \ ,
\end{equation}
where, $\lambda^{a}$ are the Gell-Mann matrices. The field $s_{0}$ ($\cal A$) represents the CP-even (-odd) gauge-singlet scalar component, and $G_{H}^{a}$ is a set of scalar color-octets. As mentioned, the CP-even scalar degree of freedom, $s_{0}$, develops a VEV, $v_{s}$, triggering spontaneous symmetry breaking in the extended color sector. The states $G_{G}^{a}$ denote the colored Nambu-Goldstone bosons, `eaten' by the colorons as a result of the symmetry breaking. The entire $\Phi$ field transforms as the bi-fundamental of the $SU(3)_{1c}\times SU(3)_{2c}$ gauge group
\begin{equation}
\Phi\to u_1 \Phi \, u_2^\dagger  \qquad \pbrac{u_i=\exp \tbrac{i \alpha_i^a t^a}} \ ,
\end{equation}
with $\alpha_i^a$ the parameters of the original $SU(3)_{ic}$ transformations. Thus, the color covariant derivative takes the form
\begin{equation}\label{Colcoder}
D_\mu \Phi = \partial_\mu \Phi - i g_{s_1} G^a_{1\mu} t^a \Phi + i g_{s_2} \Phi\,  G^a_{2\mu} t^a \ .
\end{equation}

The most general renormalizable scalar potential \cite{Hill:1993hs,Dicus:1994sw,Chivukula:1996yr,Bai:2010dj}, also formally accommodating a mixing between the $\phi$~and~$\Phi$ fields, can be written as\footnote{We follow the analysis givin in \protect\cite{Bai:2010dj}, with some modifications. See Appendix~\ref{potanal} for details.}
\begin{equation}\label{pot}
\begin{split}
V(\phi,\Phi) = &\, \frac{\lambda_s}{6}\pbrac{{\rm Tr}\tbrac{\Phi^\dagger \Phi}}^2 + \frac{\kappa_s}{2} {\rm Tr}\tbrac{\pbrac{\Phi^\dagger \Phi}^{2}} -\frac{\lambda_s + \kappa_s}{\sqrt{6}}\,r_\Delta v_{s} \pbrac{{\rm det}\Phi + {\rm h.c.}}  -\frac{\lambda_s + \kappa_s}{6} \, v_s^{2} \pbrac{1 - r_{\Delta}} {\rm Tr}\tbrac{\Phi^\dagger \Phi} \\
& +\frac{\lambda_h}{6}\pbrac{\phi^\dagger \phi - \frac{v^2_h}{2}}^2 + \lambda_m\pbrac{\phi^\dagger \phi - \frac{v^2_h}{2}} \pbrac{{\rm Tr}\tbrac{\Phi^\dagger \Phi} - \frac{v^2_s}{2}} \ ,
\end{split}
\end{equation}
where, $\lambda_{h}$, $\lambda_{m}$, $\lambda_{s}$, $\kappa_{s}$, and $r_{\Delta}$ are all dimensionless couplings.  Defining
\begin{equation}\label{lambdasp}
\lambda_{s}^{\prime} \equiv \lambda_{s} + \kappa_{s} \ ,
\end{equation}
the potential \eqref{pot} is bounded from below for large field values once the following conditions are satisfied
\begin{equation}\label{stab}
\lambda_h > 0 \ , \qquad \lambda_s^{\prime} > 0 \ , \qquad \kappa_{s} > 0 \ , \qquad \lambda_m^2 < \frac{1}{9} \lambda_h \lambda_s^{\prime} \ .
\end{equation}
Moreover, as explained in Appendix~\ref{potanal} the potential in \eqref{pot} has a global minimum for the VEVs 
\begin{equation}\label{VEVs}
\langle \phi \rangle = \frac{v_h}{\sqrt{2}} \begin{pmatrix} 0 \\ 1\end{pmatrix} \ , \qquad \langle \Phi \rangle = \frac{v_s}{\sqrt{6}} \,\mathcal I_{3\times 3} \ ,
\end{equation}
where we take $v_{h,s}>0$ by convention, provided that
\begin{equation}\label{rdel}
0 \le r_{\Delta} \le \frac{3}{2} \ .
\end{equation}

In the broken symmetry phase in both the electroweak and the extended color sectors, the kinetic terms of the quadratic Lagrangian are diagonal. However, there is now a mass-mixing among the $G^{a}_{1 \mu}$ and $G^{a}_{2 \mu}$ vector fields, and among the $h_{0}$ and $s_{0}$ scalars. These may be diagonalized by means of orthogonal rotations, which define their corresponding mass eigenstates
\begin{equation}\label{massbasis}
\begin{pmatrix} G^{a}_{1 \mu}\\ G^{a}_{2 \mu} \end{pmatrix}
= R_{\theta_{c}} \begin{pmatrix} G^{a}_{\mu} \\ C^{a}_{\mu} \end{pmatrix} \ , \qquad
\begin{pmatrix} h_0\\ s_0 \end{pmatrix}
= R_{\chi} \begin{pmatrix} h \\ s \end{pmatrix} \ ,
\end{equation}
with the mixing angles $\theta_{c}$ and $\chi$, respectively, and 
\begin{align}
R_{\theta_{c}}\equiv &\, \begin{pmatrix} \cos\theta_{c} & -\sin\theta_{c} \\ \sin\theta_{c} & \cos\theta_{c} \end{pmatrix} \ , \qquad
\sin\theta_c \equiv \frac{g_{s_1}}{\sqrt{g_{s_1}^2+g_{s_2}^2}} \ , \label{Rtheta} \\
R_{\chi}\equiv &\, \begin{pmatrix} \cos\chi & \sin\chi \\ -\sin\chi & \cos\chi \end{pmatrix} \ , \qquad
\cot 2\chi \equiv \frac{1}{6\lambda_m} \tbrac{ \lambda_s^{\prime}\pbrac{1-\frac{r_{\Delta}}{2}} \frac{v_s}{v_h} - \lambda_h \frac{v_h}{v_s} } \label{Rchi}\ .
\end{align}
The gluon field, $G^{a}_{\mu}$ (c.f. \eqref{massbasis}), remains massless, with the corresponding QCD coupling, $g_{s}$, defined by
\begin{equation}\label{gs}
\frac{1}{g_s^2} = \frac{1}{g_{s_1}^2}+\frac{1}{g_{s_2}^2}\ ,
\end{equation}
whereas, the coloron, $C^{a}_{\mu}$, acquires the mass \cite{Bai:2010dj}
\begin{equation}\label{MC}
M_C = \sqrt{\frac{2}{3}}\frac{g_s\, v_{s}}{\sin 2\theta_c} \ .
\end{equation}
The masses of the diagonalized scalar states, $h$ and $s$ in \eqref{massbasis}, are determined to be
\begin{equation}\label{mhs}
m_{h,s}^{2} = \frac{1}{6} \cbrac{ \lambda_h v_h^2 + \lambda_s^{\prime} v_s^2\pbrac{1-\frac{r_{\Delta}}{2}} \pm \tbrac{\lambda_h v_h^2 - \lambda_s^{\prime} v_s^2\pbrac{1-\frac{r_{\Delta}}{2}}}\sec 2\chi } \ .
\end{equation}

In this paper, we shall identify the lightest of these two massive scalar degrees of freedom, $h$, with the recently discovered 125 GeV Higgs-like state at the LHC \cite{Aad:2012tfa,Chatrchyan:2012ufa}. Furthermore, from the quadratic Lagrangian, one can deduce the following (already diagonal) masses for the other physical (pseudo-)scalars in the spectrum
\begin{align}
m_{\cal A}^{2} =&\, \frac{v_{s}^{2} }{2} \, r_{\Delta} \lambda_{s}^{\prime} \ , \label{mA}\\
m_{G_{H}}^{2} =&\, \frac{1}{3} \tbrac{v_{s}^{2} \, \kappa_{s}+ 2 m_{\cal A}^{2}} \ , \label{mGH}
\end{align}
which, invoking \eqref{stab}, implies the condition
\begin{equation}\label{masscond}
m^2_{G_{H}} \geq \frac{2}{3} m^2_{\mathcal A} \ .
\end{equation}

\subsection{The Fermionic Sector}
\label{subsec:fermions}

Generally speaking, the charges of the ordinary quarks under the extended $SU(3)_{1c} \times SU(3)_{2c}$ color gauge group can be either vectorial or chiral. Regardless of how we choose those charges, in the vector boson mass basis \eqref{massbasis} the quarks will display their usual vectorial coupling to the gluon; their coupling to the massive coloron, however, may be chiral \cite{Chivukula:2011ng, Chivukula:2013xla}. If the ordinary fermions have chiral charges under the extended color group, this will render
the $SU(3)_{1c} \times SU(3)_{2c}$ theory anomalous 
\cite{Frampton:1987dn,Frampton:1987ut,Cvetic:2012kv,Chivukula:2013xla}.
As described in \cite{Cvetic:2012kv}, the simplest way to cancel these anomalies is to include additional spectator fermions, $Q^{k}_{L,R}$ which simultaneously (a) have the same electric charges as the ordinary quarks, (b) have the opposite chirality charges as the ordinary quarks under $SU(3)_{1c} \times SU(3)_{2c}$, and (c) are vectorial under the $SU(2)_W \times U(1)_Y$ electroweak interactions (e.g., are all weak doublets or all weak singlets).\footnote{In general, if the spectators were, instead, chosen to be chiral under the electroweak interactions, additional
lepton-like (color neutral) spectators would be required to cancel $SU(2)_W$ global anomalies.} Here $Q^k_{L,R}$ denote up- and down-type quarks and $k$ is a flavor index. Since the spectators are vectorial under the electroweak interactions they can get a mass from a Yukawa coupling with the $\Phi$ boson. The fermion Lagrangian then contains
\begin{equation}\label{Lferm}
{\cal L}_{\rm fermion} \supset - y_{Q} \tbrac{\bar{Q}^{k}_{R} \, \Phi\, Q^{k}_{L} + \bar{Q}^{k}_{L} \, \Phi^\dagger \, Q^{k}_{R}} \ ,
\end{equation}
where, for simplicity, the spectator fermion masses are assumed to be flavor-universal
\begin{equation}\label{MQ}
M_{Q} = \frac{y_{Q}}{\sqrt 6}\, v_{s} \ .
\end{equation}
Due to the strong constraints on flavor-changing couplings of the colorons, we neglect the potential mixing between the ordinary quarks and the spectator fermions \cite{Chivukula:2013kw}.

The number of spectator fermions required depends on the details of the model \cite{Cvetic:2012kv}. If the ordinary quarks' charges are vectorial under the extended color interactions, no spectators are necessary. If the quarks' extended color interactions are chiral but flavor-universal, then three generations of spectators (three up-like and three down-like spectators) are required. Alternatively, if one takes the chiral couplings of the third quark generation under the extended color group to be opposite to those of the first two, then only one spectator generation (one up-like and one down-like spectator) is needed to cancel anomalies. In what follows, therefore, we will present phenomenological results (see section \ref{LHCsearch}) in the case of 0, 1, or 3 generations of spectator quarks.

The relevant Feynman rules of the renormalizable coloron model are listed in Appendix~\ref{FR}. The following sections of the current study  explores the viability of this theory by investigating the formal and phenomenological consequences of its predicted interactions.

\section{Theoretical Bounds on the Model}\label{boundsall}

In this section, we examine the current general constraints on the renormalizable coloron  model, arising from various theoretical considerations; experimental bounds will be considered in the next section. Compared to the ordinary SM, the renormalizable coloron  model introduces seven additional free parameters:
\begin{itemize}
\item Five free parameters from the enhanced scalar sector (Eq. \eqref{pot}); namely, the CP-even singlet VEV, $v_s$, and the dimensionless couplings $\kappa_{s}$, $r_{\Delta}$, $\lambda_{m}$, and $\lambda_{s}$ (or equivalently $\lambda_{s}^{\prime}$ by using \eqref{lambdasp}), 
\item One free parameter from the extended color gauge group (Eq. \eqref{gs}), which can be taken as the gauge group's mixing angle, $\theta_{c}$;
\item One free parameter from the extended fermion sector (Eq. \eqref{Lferm}); namely, the spectators' (universal) Yukawa coupling, $y_{Q}$.
\end{itemize}
With the aid of \eqref{Rchi}, \eqref{MC}-\eqref{mGH}, and \eqref{MQ}, it is possible to trade all of the scalar dimensionless couplings, the color gauge group's mixing angle, and the spectators' Yukawa coupling for the physically more convenient mass spectrum of the particles present in the theory and the scalar mixing angle, according to the definitions
\begin{equation}\label{conversions}
\begin{split}
\lambda_{h} = &\, \frac{3}{2}\frac{m_{h}^{2}+m_{s}^{2}+\pbrac{m_{h}^{2}-m_{s}^{2}}\cos 2\chi}{v_{h}^{2}} \ , \qquad \lambda_{m} = -\frac{1}{2}\frac{m_{h}^{2}-m_{s}^{2}}{v_{h} v_{s}}\sin 2\chi \ , \\
\lambda_{s}^{\prime} = &\, \frac{1}{2}\frac{2m_{\mathcal A}^{2}+3\pbrac{m_{h}^{2}+m_{s}^{2}}-3\pbrac{m_{h}^{2}-m_{s}^{2}}\cos 2\chi}{v_{s}^{2}} \ , \qquad \kappa_{s} = \frac{3m_{G_{H}}^{2}-2m_{\mathcal A}^{2}}{v_{s}^{2}} \ , \\
r_{\Delta} = &\, \frac{4 m_{\mathcal A}^{2}}{2m_{\mathcal A}^{2}+3\pbrac{m_{h}^{2}+m_{s}^{2}}-3\pbrac{m_{h}^{2}-m_{s}^{2}}\cos 2\chi} \ , \qquad \sin 2\theta_{c} = \sqrt \frac{2}{3} \frac{g_{s} \, v_{s}}{M_{C}} \ , \qquad y_{Q} =\sqrt 6 \frac{M_{Q}}{v_{s}} \ .
\end{split}
\end{equation}
Note that the sign of $\sin\chi$ is correlated with the sign of $\lambda_m$, and that either sign is possible (see Eq. \eqref{stab}).
Fixing the electroweak VEV at $v_h = 246$~GeV and the mass of the $h$ scalar at $m_{h}=125$~GeV, the seven new free parameters of the renormalizable coloron model can be conveniently written as
\begin{equation}\label{freepar}
\cbrac{v_{s}, m_{s}, \sin \chi, m_{\cal A}, m_{G_{H}}, M_{C}, M_{Q}} \ .
\end{equation}

In the following sub-sections, we constrain the model's parameter space \eqref{freepar} on several theoretical grounds. For the sake of physical clarity, we shall display the resulting bounds in two dimensional exclusion plots of $m_{s}$ vs. $\sin \chi$, for various benchmark values of the remaining relevant parameters entering each analysis. In presenting these plots, we take into account existing collider limits on the masses of the coloron and the color-octet scalar. Older Tevatron and current LHC searches require the coloron mass, $M_{C}$, to be at least in the TeV region \cite{Simmons:1996fz,Bertram:1998wf,ATLAS:2012pu,ATLAS:2012qjz,Chatrchyan:2013qha,CMS:kxa}, while the Tevatron excludes the mass range from  50 to 125 GeV for the scalar color-octet, $m_{G_{H}}$ \cite{Aaltonen:2013hya}.

\subsection{Stability} \label{stability}

As mentioned in section~\ref{intro}, the scalar potential \eqref{pot} is guaranteed to be bounded from below for large values of the fields, once the conditions \eqref{stab} are fulfilled. Trading the relevant couplings for the model's free parameters \eqref{freepar} by means of \eqref{conversions}, one can, however, easily demonstrate that all of these conditions are automatically satisfied in the full $m_{s}-\sin \chi$ parameter space.\footnote{It is curious to note that, given \eqref{conversions}, the coupling $\kappa_{s}$ bears no relation to $m_{s}$ and $\sin \chi$, and, hence, its stability condition \eqref{stab} is trivially satisfied.}

\begin{figure}
\begin{center}
\includegraphics[width=.329\textwidth]{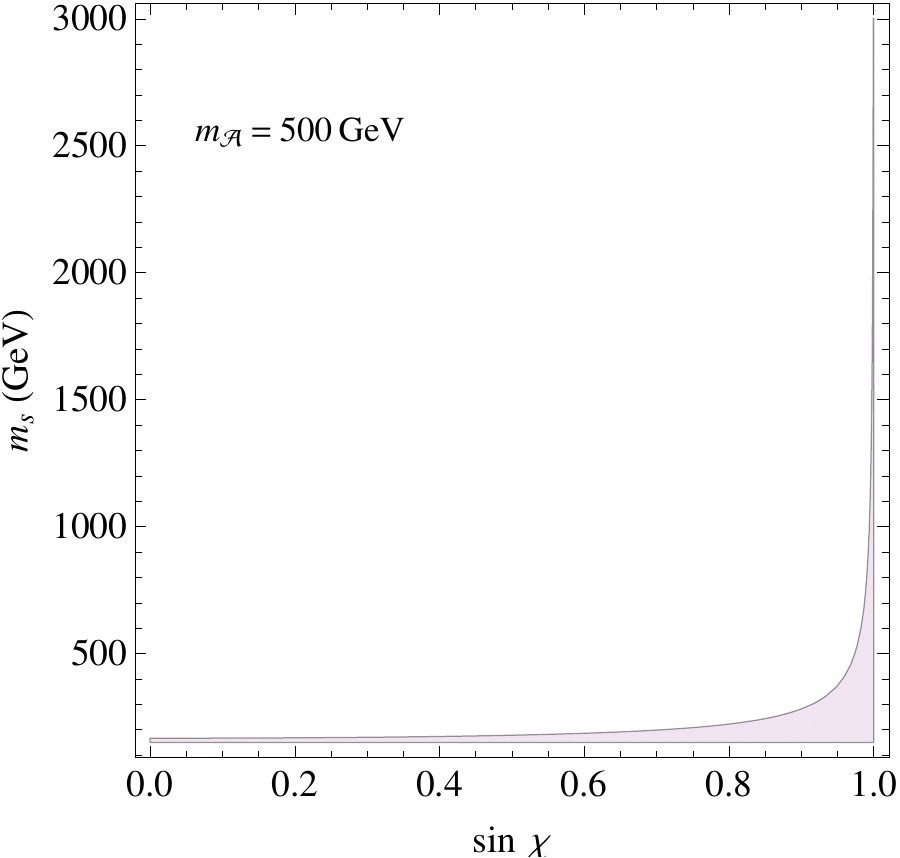}
\includegraphics[width=.329\textwidth]{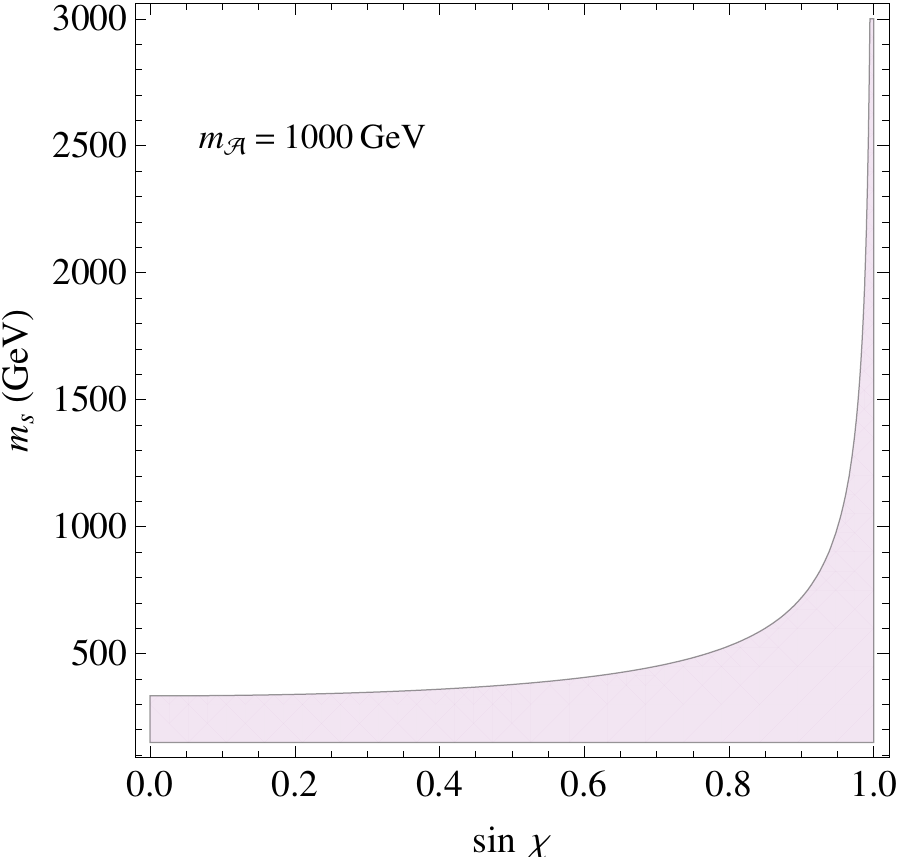}
\includegraphics[width=.329\textwidth]{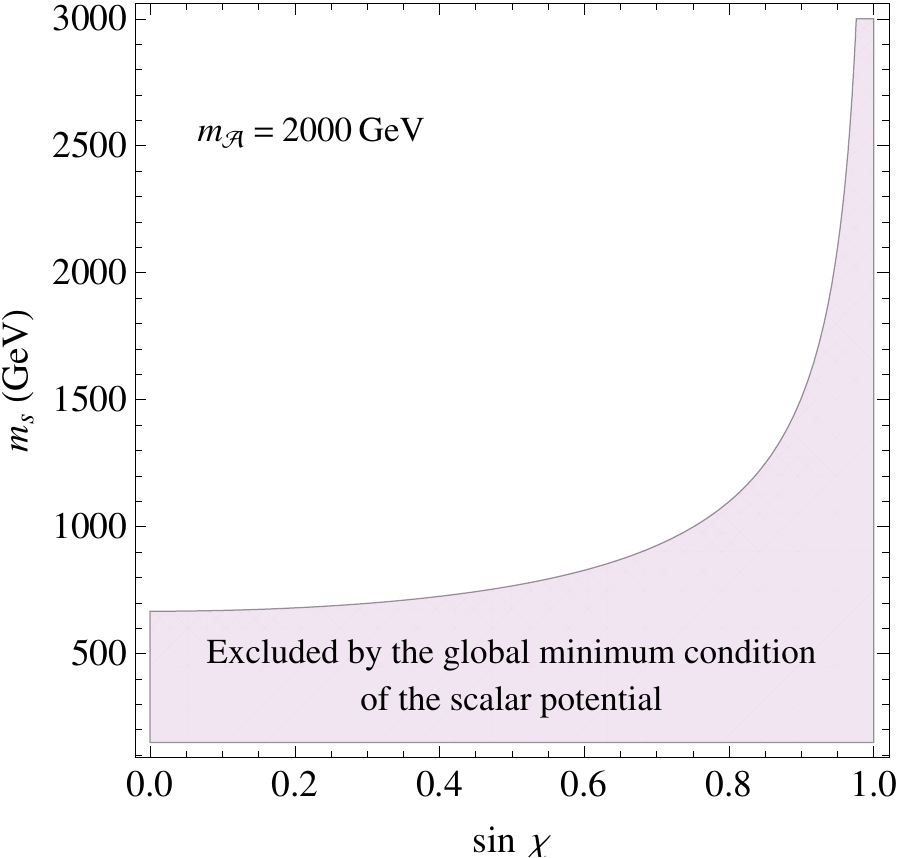}
\caption{Constraints on the $m_{s}-\sin \chi$ plane arising from the scalar potential's global minimum condition \eqref{rdel}, for $m_s\geq 150$~GeV. The panels correspond to three representative values of the $\mathcal A$ pseudo-scalar mass (c.f. \eqref{Phi}). A heavy $m_{\mathcal A}$ generally disfavors a light $m_{s}$ for any scalar mixing angle value.}
\label{stabmA}
\end{center}
\end{figure}

The only non-trivial constraint arises from the condition on $r_{\Delta}$ in Eq. \eqref{rdel}, which was necessary to ensure that the global minimum of the scalar potential would coincide with the electroweak and the extended color sector symmetry breaking VEVs \eqref{VEVs}. Inspecting \eqref{conversions} reveals that the condition \eqref{rdel} on $r_{\Delta}$ translates into bounds in the $m_{s}-\sin \chi$ plane for various $m_{\mathcal A}$ values, and is independent of the other free parameters of the theory. In particular, an explicit dependence on the singlet VEV, $v_{s}$, cancels. Fig.~\ref{stabmA} depicts the exclusion bounds on the $m_{s}-\sin \chi$ plane, arising from the condition \eqref{rdel}, for several benchmark values of $m_{\mathcal A}$. While a light $m_{\mathcal A}$ leaves the plane unconstrained, it is evident that a heavy $\mathcal A$ pseudo-scalar disfavors a relatively light $s$ scalar for all values of the scalar mixing angle.
In particular, examining the expression for $r_\Delta$ in \eqref{conversions}, we see that in the limit\footnote{As we demonstrate in sec. \protect\ref{sec:EWPT} below, the precision electroweak data implies that $\sin\chi$ is less than about 0.2.} $\sin \chi \to 0$
\begin{equation}
m_s \ge  \frac{m_{\mathcal A}}{3}~.
\end{equation}

\subsection{Unitarity} \label{unitarity}

\begin{figure}
\begin{center}
\includegraphics[width=.55\textwidth]{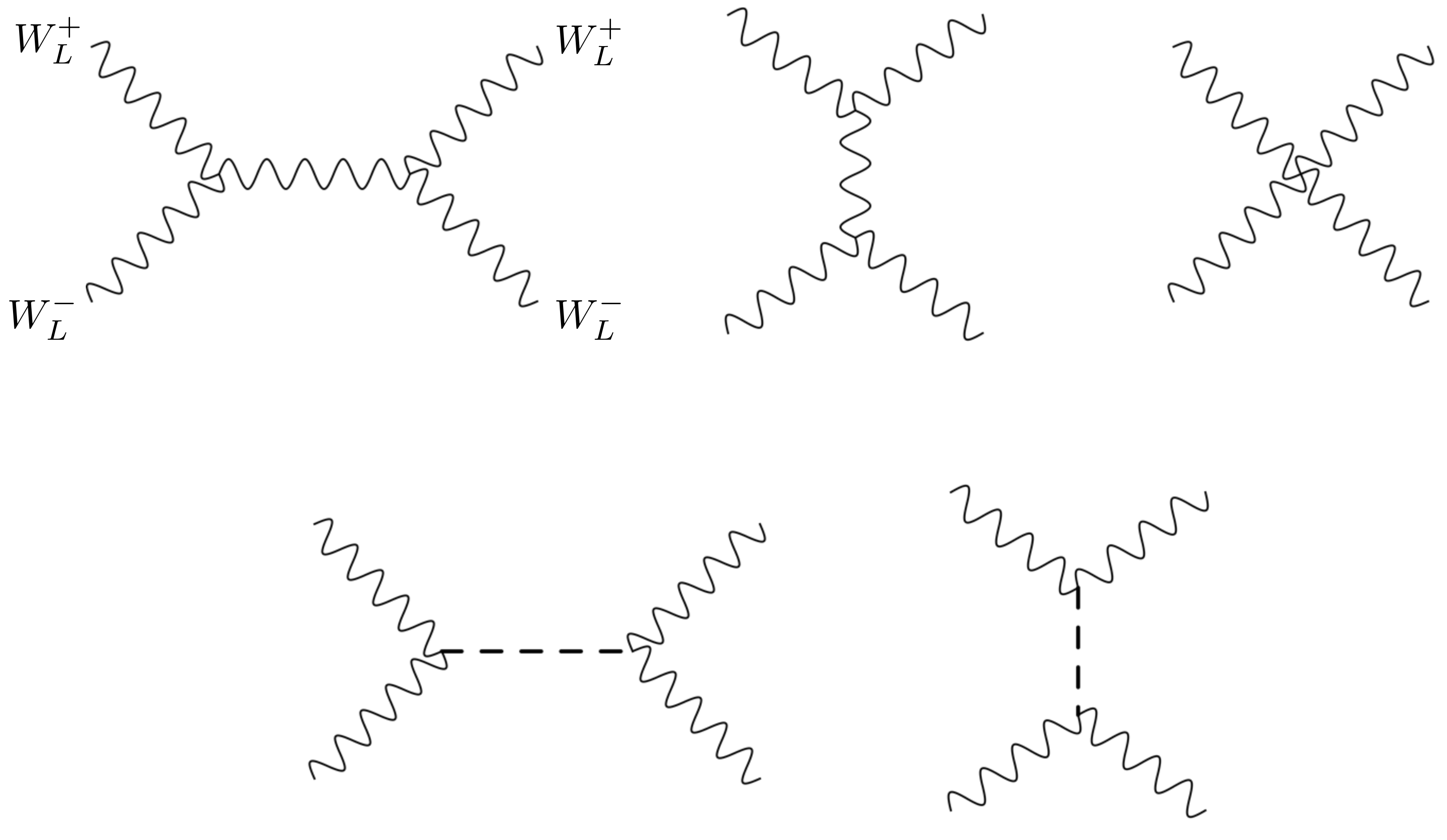}
\caption{Diagrams contributing to the $WW$ scattering. The top row illustrates the $s$- and $t$-channels of the electroweak gauge boson exchange, as well as the four-point contact interaction. The bottom row diagrams are due to scalar exchange.}
\label{Unit}
\end{center}
\end{figure}

It is well-established, within the SM, that the high energy cross section for the tree-level $W_L^+ W_L^- \to W_L^+ W_L^-$ scattering is unitarized by the  exchange of the SM Higgs boson in the $s$- and $t$-channels. To be specific, the Higgs boson exchange contributions cancel the term of the cross-section that is linear in the center of mass  (CM) energy, introduced by the electroweak gauge boson exchange channels as well as the $W$ four-point contact interaction (Fig.~\ref{Unit}). The tree-level cross section, consequently, depends only logarithmically on the CM energy and respects unitarity to very high energies by satisfying the condition
\begin{equation}\label{unitcon}
\left| a_0 \right| < \frac{1}{2} \ .
\end{equation}
In this expression, $a_0$ is the s-wave coefficient of the partial-wave expanded amplitude, given by
\begin{equation}\label{a0}
a_0 = \frac{1}{32\pi} \int_{-1}^1 \mathcal{T}\,d\cos\theta \ ,
\end{equation}
where $i\mathcal{T}$ is the total amplitude of the scattering process, and $\theta$ is the polar angle in spherical coordinates.

In the renormalizable coloron model, the unitarizing role of the SM Higgs, $h_{0}$, is divided between the two scalars $h$ and $s$, according to the mixing \eqref{massbasis}. The corresponding couplings are suppressed with respect to the original SM Higgs coupling by factors of $\cos \chi$ and $\sin \chi$, respectively. The resulting total scalar contribution to the linear CM energy term in the cross section is, thus, proportional to $\cos^{2} \chi + \sin^{2} \chi = 1$, and the same cancellation against the gauge boson diagrams follows as in the SM. The high energy cross section for tree-level $W_L^+ W_L^- \to W_L^+ W_L^-$ scattering is, consequently, only logarithmically dependent on the CM energy, as in the ordinary SM.

Let us now turn to a general study of the unitarity of longitudinal electroweak vector boson scattering in this model. The unitarity analysis concerns, in particular, the high energy behavior of these cross sections, with the CM energy exceeding any other mass scale present in the theory. In this energy regime, the external longitudinal vector bosons may be well-approximated by their corresponding eaten Nambu-Goldstone bosons, according to the equivalence theorem \cite{Dicus:1992vj,Cornwall:1974km,Vayonakis:1976vz,Veltman:1976rt,Lee:1977eg,Chanowitz:1985hj,Gounaris:1986cr}.   A full coupled-channel analysis of the scalar scattering processes involving all the neutral initial- and final-state two-body scalar scatterings \cite{Lee:1977eg} is appropriate. Furthermore, one may perform the unitarity study in the gaugeless limit, where all the gauge couplings are set equal to zero and the contribution of the gauge boson exchange channels may be neglected.

The overall strategy can be summarized as follows: the coupled-channel amplitude of all the neutral initial- and final-state two-body scalar scatterings in the theory can be written as
\begin{equation}\label{amp}
i\mathcal{T}_{\text{coupled}} = b \cdot i T \cdot b^T \ ,
\end{equation}
where $b$ is the normalized basis of the initial two-body asymptotic states, and $iT$ is the scattering matrix connecting the initial and final asymptotic states. Once the coupled-channel amplitude \eqref{amp} is determined, it may be expanded in partial waves by means of \eqref{a0}. In order to deduce the most stringent constraints on the parameter space \eqref{freepar} that are dictated by the unitarity condition \eqref{unitcon}, one needs to diagonalize the resulting expression and find its largest eigenvalue(s) \cite{Lee:1977eg}.

Employing the field eigenstates,\footnote{In principle, the amplitude \eqref{amp} needs to be evaluated in the mass eigenstate basis.  However,  the amplitudes constructed in the mass and in the field eigenstate bases share the same eigenvalues. Since we are ultimately interested in the eigenvalues of the partial-wave expanded coupled-channel amplitude, we choose to work in the (computationally more convenient) field eigenstate.} the properly normalized and (color and electric) neutral scattering basis, $b$, takes the form
\begin{align}
b =&\, \begin{pmatrix} b_{1} & b_{2} & b_{3} \end{pmatrix} \ , \label{b} \\
b_{1} \equiv&\, \begin{pmatrix} \pi^+ \pi^- & \frac{1}{\sqrt 2} h_0h_0 & \frac{1}{\sqrt 2} \pi^0\pi^0 & \frac{1}{\sqrt 2} s_0s_0 & \frac{1}{\sqrt 2} \cal A\cal A & \frac{1}{4} G_{G}^{a} G_{G}^{a}& \frac{1}{4} G_{H}^{a}G_{H}^{a} \end{pmatrix} \ , \label{b1} \\
b_{2} \equiv&\, \begin{pmatrix} \frac{1}{\sqrt 8} G_{G}^{a}G_{H}^{a} & s_0 \cal A \end{pmatrix} \ , \label{b2} \\
b_{3} =&\, \begin{pmatrix} h_0s_0 & \pi^{0}s_0 & h_0 \cal{A} & \pi^{0}\cal{A} & h_0\pi^0  \end{pmatrix} \ , \label{b3}
\end{align}
where implicit summation over the color factors is assumed. As we shall demonstrate, the scattering matrix, $iT$, in \eqref{amp} is of a block-diagonal form, with each block corresponding to a sub-basis $b_{i}$, given in \eqref{b1}-\eqref{b3}.

In the high CM energy limit, the scalar exchange channels are suppressed compared to the scalar contact interactions, due to the intermediate propagator which falls off with the energy. Because we are working in the gaugeless limit, there are no vector boson exchange diagrams. Consequently, for the purpose of analyzing the unitarity constraints in this energy regime, it is sufficient to take into account only the scalar contact diagrams.

Using the normalized neutral scattering basis \eqref{b} and the Feynman rules listed in Appendix~\ref{FR}, one can determine the scattering matrix, $iT$, from the scalar quartic couplings. We find it takes block-diagonal form, when written using the sub-bases \eqref{b1}-\eqref{b3}, and is given by
\begin{equation}\label{T}
T = - \begin{pmatrix}
d_{1} &0 & 0 \\
0 & d_{2} & 0 \\
0 & 0 & d_{3}\\
\end{pmatrix} \ ,
\end{equation}
with the blocks defined by
\begin{equation}\label{ai}
\begin{split}
d_{1} =& \begin{pmatrix}
\frac{2\lambda_h}{3} & \frac{\lambda_h}{3\sqrt 2} & \frac{\lambda_h}{3\sqrt 2} & \frac{\lambda_m}{\sqrt 2} & \frac{\lambda_m}{\sqrt 2} & 2\lambda_m & 2\lambda_m \\
 \frac{\lambda_h}{3\sqrt 2} & \frac{\lambda_h}{2} &\frac{\lambda_h}{6} & \frac{\lambda_m}{2} & \frac{\lambda_m}{2} & \sqrt2\lambda_m & \sqrt2\lambda_m \\
 \frac{\lambda_h}{3\sqrt 2}  &\frac{\lambda_h}{6} &\frac{\lambda_h}{2} & \frac{\lambda_m}{2} & \frac{\lambda_m}{2} & \sqrt2\lambda_m & \sqrt2\lambda_m \\
\frac{\lambda_m}{\sqrt 2} & \frac{\lambda_m}{2} & \frac{\lambda_m}{2} & \frac{\lambda_s^{\prime}}{2} &  \frac{\lambda_s^{\prime}}{6} & \frac{\sqrt2\lambda_s^{\prime}}{3} & \sqrt2\frac{\lambda_s^{\prime}+2 \kappa_{s}}{3} \\
\frac{\lambda_m}{\sqrt 2} & \frac{\lambda_m}{2} & \frac{\lambda_m}{2} & \frac{\lambda_s^{\prime}}{6} & \frac{\lambda_s^{\prime}}{2} & \sqrt2\frac{\lambda_s^{\prime}+2 \kappa_{s}}{3} & \frac{\sqrt2\lambda_s^{\prime}}{3} \\
2\lambda_m & \sqrt2\lambda_m & \sqrt2\lambda_m & \frac{\sqrt2\lambda_s^{\prime}}{3} & \sqrt2\frac{\lambda_s^{\prime}+2 \kappa_{s}}{3}& 5\frac{2\lambda_s^{\prime}+ \kappa_{s}}{6} & \frac{8\lambda_s^{\prime}+9 \kappa_{s}}{6} \\
2\lambda_m & \sqrt2\lambda_m & \sqrt2\lambda_m & \sqrt2\frac{\lambda_s^{\prime}+2 \kappa_{s}}{3} & \frac{\sqrt2\lambda_s^{\prime}}{3} & \frac{8\lambda_s^{\prime}+9 \kappa_{s}}{6} & 5\frac{2\lambda_s^{\prime}+ \kappa_{s}}{6} \\
\end{pmatrix} \ , \\
d_{2} =& \begin{pmatrix}
 \frac{\lambda_s^{\prime}-2 \kappa_{s}}{3} &\frac{\sqrt 8 \kappa_{s}}{3} \\
\frac{\sqrt 8 \kappa_{s}}{3} & \frac{\lambda_{s}^{\prime}}{3} \\
\end{pmatrix} \ , \qquad \qquad \qquad \qquad
d_{3} = \begin{pmatrix}
\lambda_{m}\mathcal I_{4\times4} &0 \\
0 & \frac{\lambda_{h}}{3}  \\
\end{pmatrix} \ .
\end{split}
\end{equation}
Inserting \eqref{T} into \eqref{amp}, and performing the trivial integration in \eqref{a0}, we obtain for the partial-wave expanded scattering amplitude in the field basis
\begin{equation}\label{a0fin}
a_{0\, \text{coupled}} = -\frac{1}{16\pi} \,T \ ,
\end{equation}
Subsequent diagonalization of $T$ \eqref{T} yields for \eqref{a0fin} the eigenvalues
\begin{equation}\label{eigenval}
\setcounter{MaxMatrixCols}{14}
a_{0\, \text{coupled}} = -\frac{1}{16\pi} \,\text{diag}\begin{pmatrix} \frac{\lambda_h}{3} & \frac{\lambda_h}{3} & \frac{\lambda_{s}^{\prime}-\kappa_{s}}{3} & \frac{\lambda_{s}^{\prime}+2\kappa_{s}}{3} & \frac{\lambda_{s}^{\prime}-4\kappa_{s}}{3} & \lambda^{-} & \lambda^{+} & \frac{\lambda_{s}^{\prime}+2\kappa_{s}}{3} & \frac{\lambda_{s}^{\prime}-4\kappa_{s}}{3} & \lambda_m & \lambda_m & \lambda_m & \lambda_m & \frac{\lambda_h}{3} \end{pmatrix} \ ,
\end{equation}
where
\begin{equation}\label{lambda+-}
\lambda^{\pm} \equiv \frac{1}{6} \left[3\lambda_h+10 \lambda_s^\prime+8 \kappa_s\pm \sqrt{648\lambda_{m}^{2} + \pbrac{3\lambda_h-10 \lambda_s^\prime-8 \kappa_s}^{2}}\right] \ .
\end{equation}
The largest eigenvalue is $\lambda^+$, and the corresponding
eigenvector is essentially the color-neutral channel composed of the colored scalars $G^a_G G^a_G$ and $G^a_H G^a_H$; scattering in this channel is enhanced by the large number of states.

The eigenvalues \eqref{eigenval} are, by virtue of \eqref{conversions}, functions of the singlet VEV, the scalar mixing angle, and the three scalar masses. Demanding the unitarity condition \eqref{unitcon} to be satisfied for each eigenvalue, we find that the most stringent constraint in the $m_{s}-\sin \chi$ plane is set by the eigenvalue $\lambda^{+}$ \eqref{lambda+-} for various (reasonable) choices of the remaining relevant input parameters. It is worth noting that, for each selected singlet VEV, $v_{s}$, the bounds are only moderately sensitive to the pseudo-scalar mass, $m_{\mathcal A}$; in contrast, their dependence on the mass of the heavier (Eq. \eqref{masscond}) scalar color-octet, $m_{G_{H}}$, is significant.

\begin{figure}
\includegraphics[width=.329\textwidth]{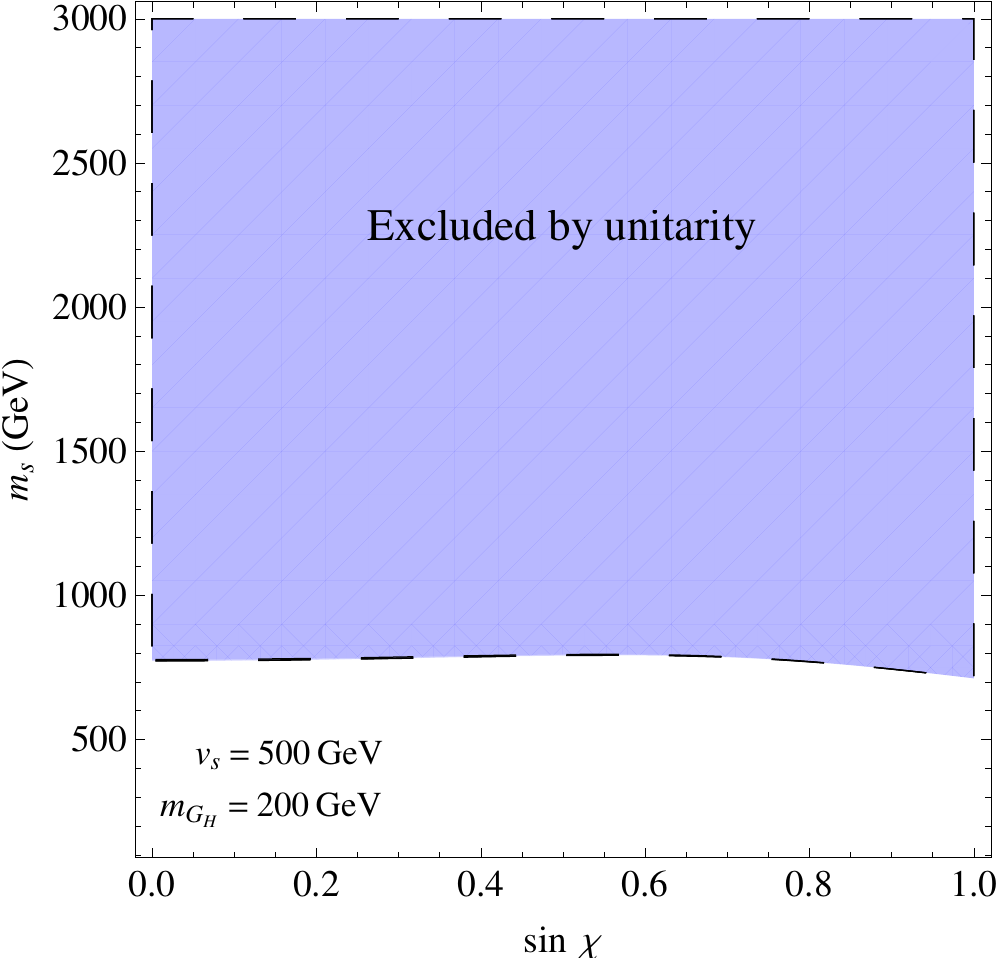}
\includegraphics[width=.329\textwidth]{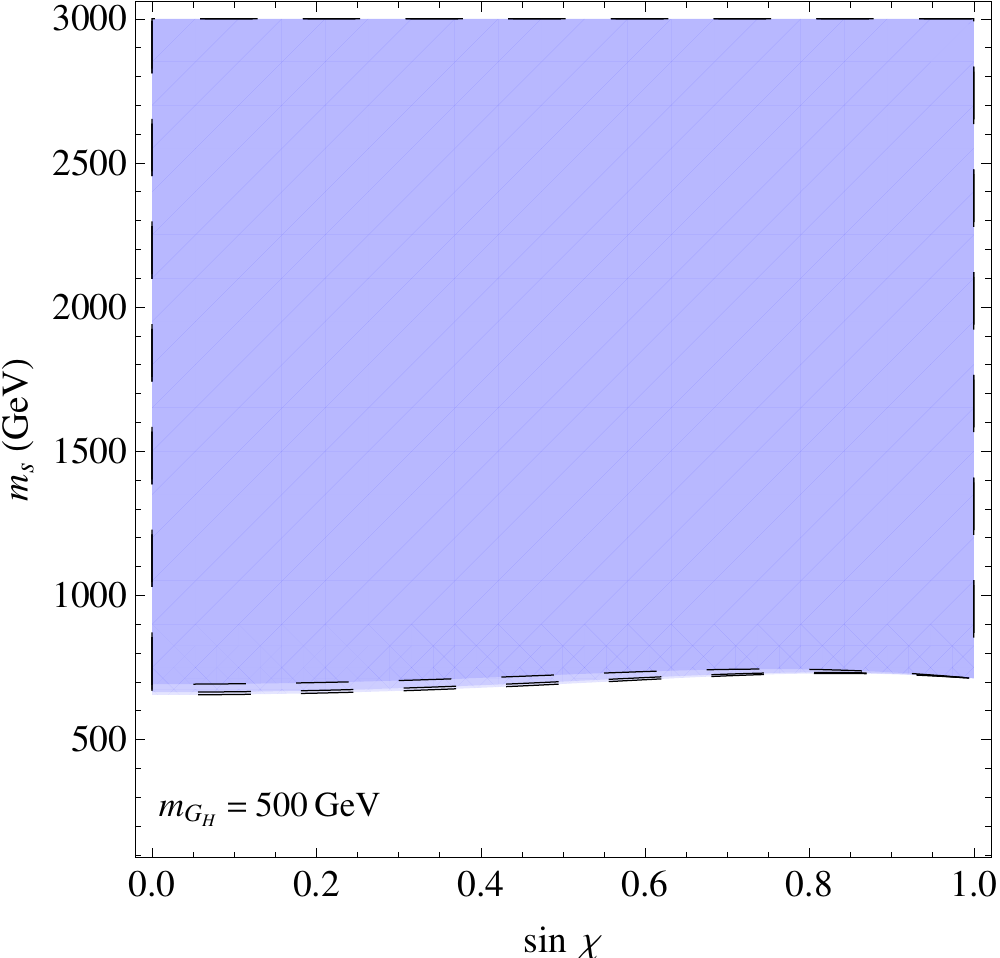}
\includegraphics[width=.329\textwidth]{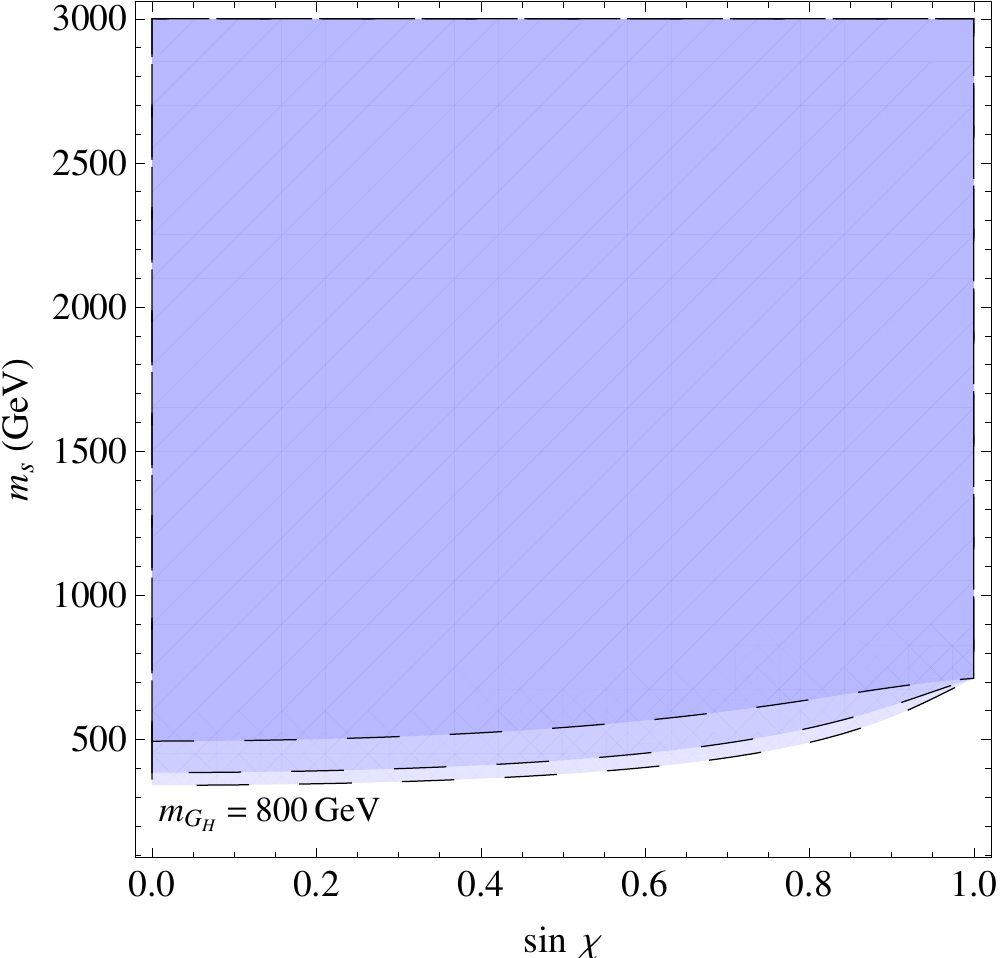}
\includegraphics[width=.329\textwidth]{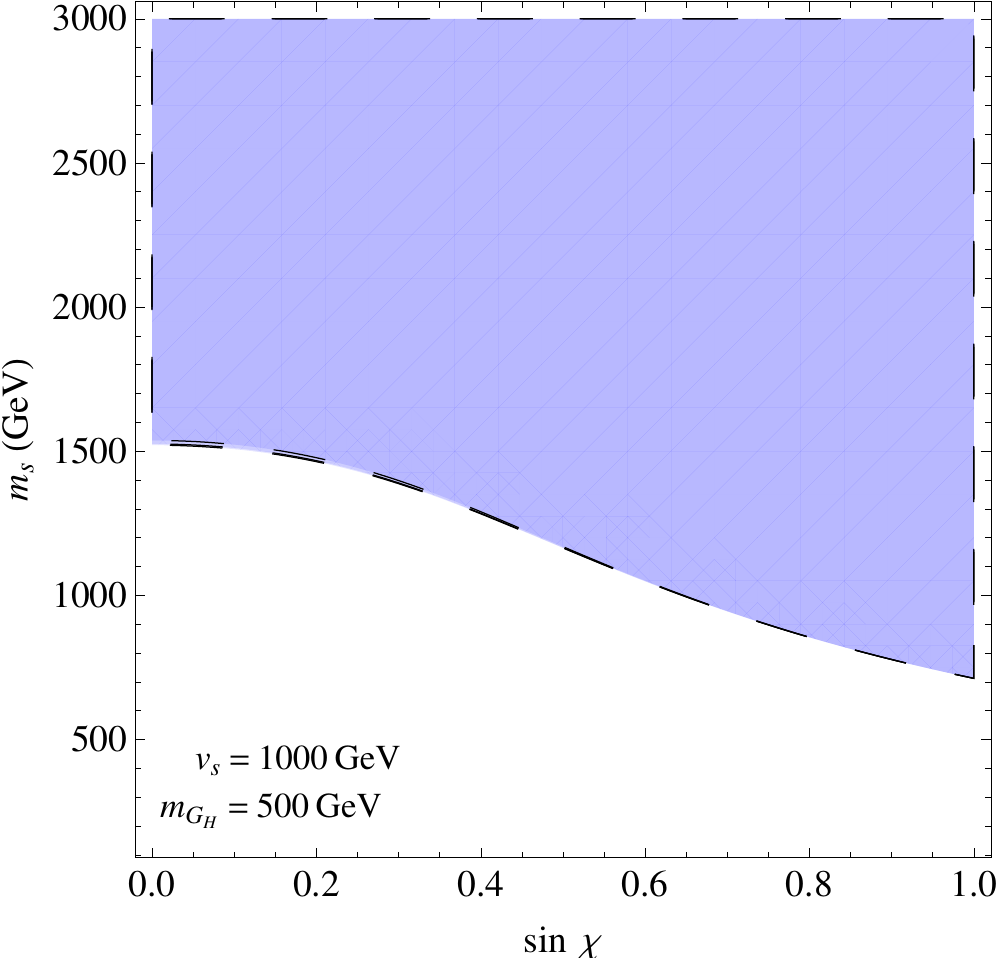}
\includegraphics[width=.329\textwidth]{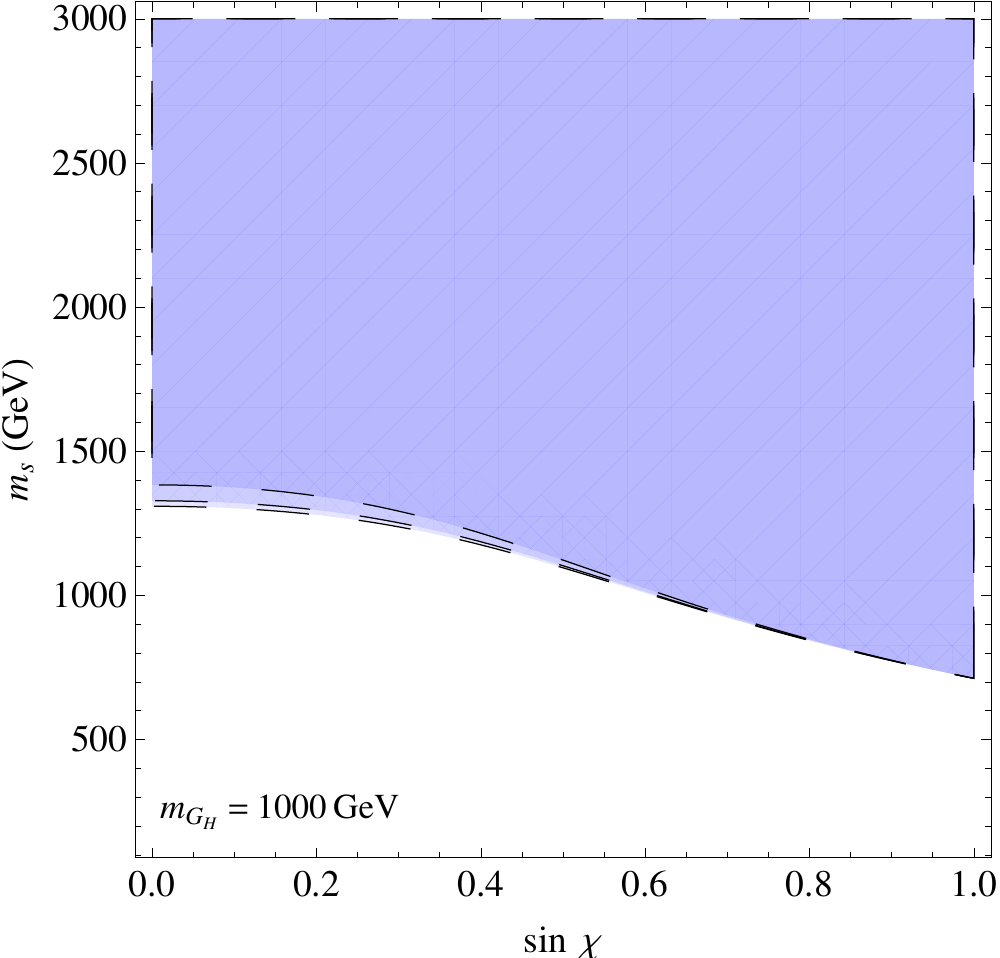}
\includegraphics[width=.329\textwidth]{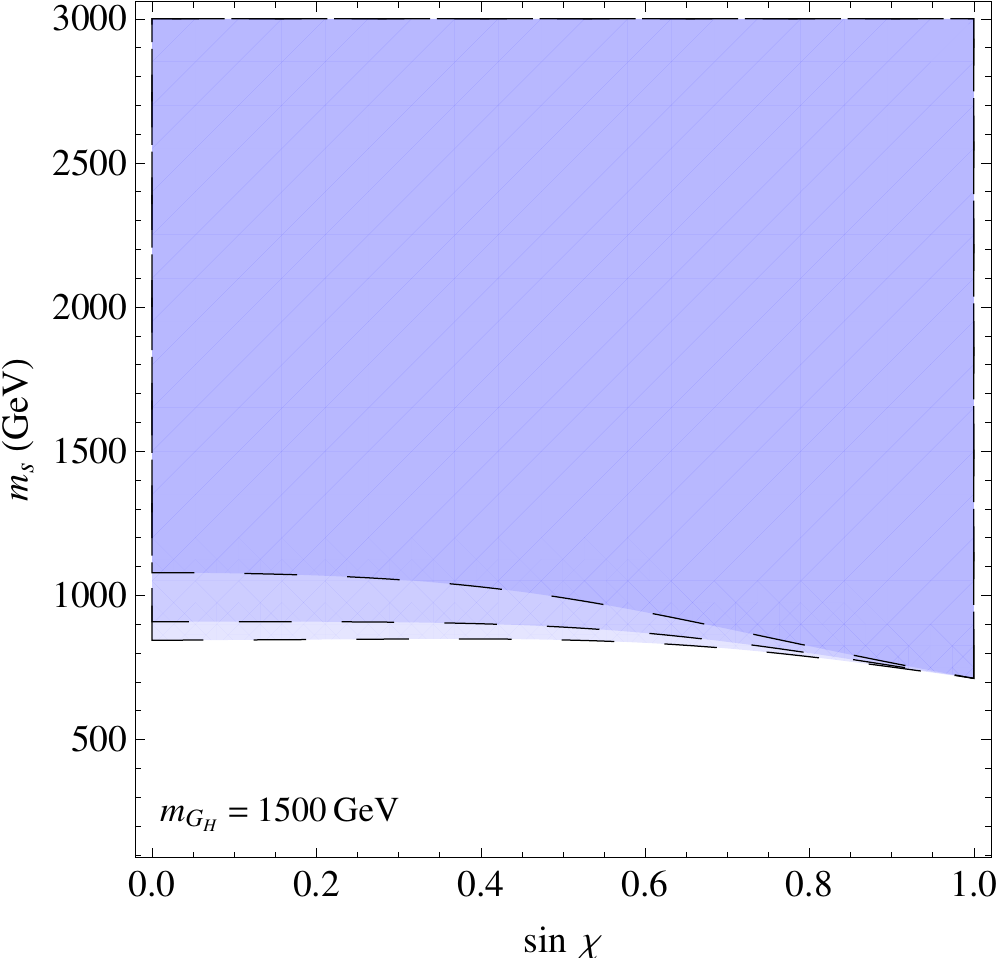}
\includegraphics[width=.329\textwidth]{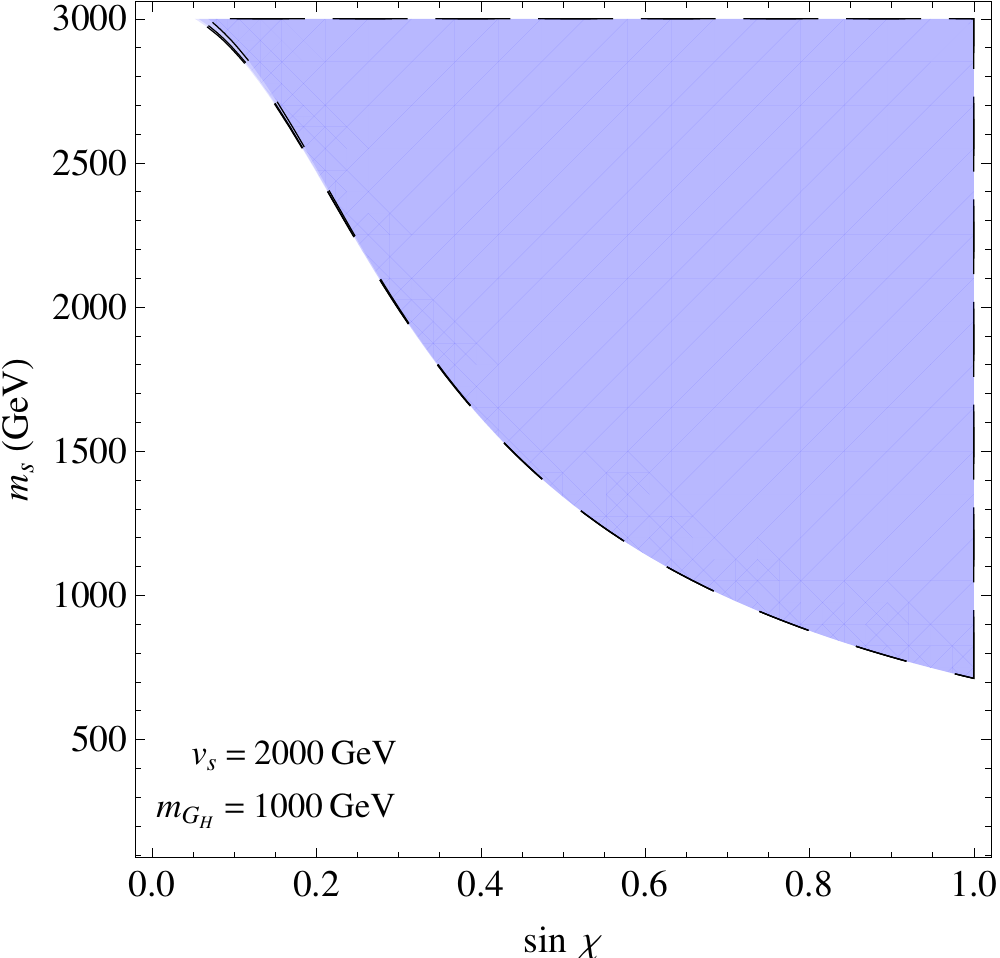}
\includegraphics[width=.329\textwidth]{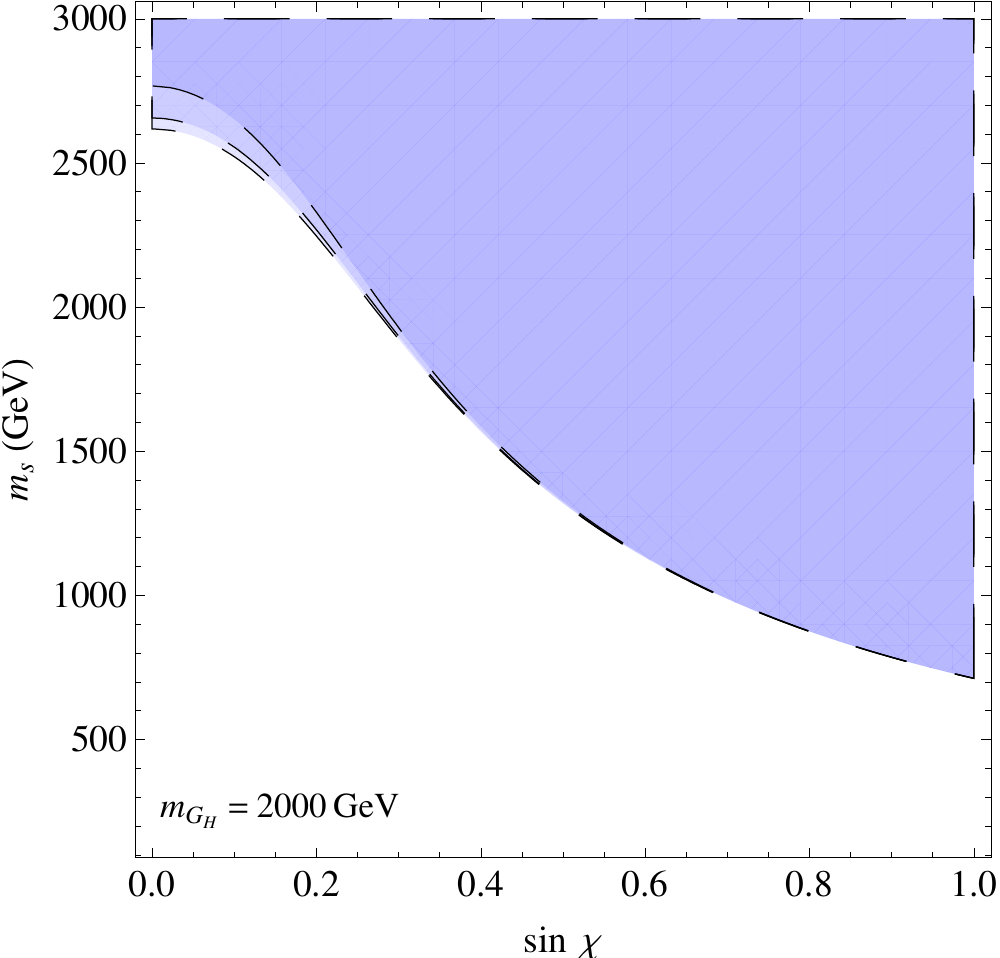}
\includegraphics[width=.329\textwidth]{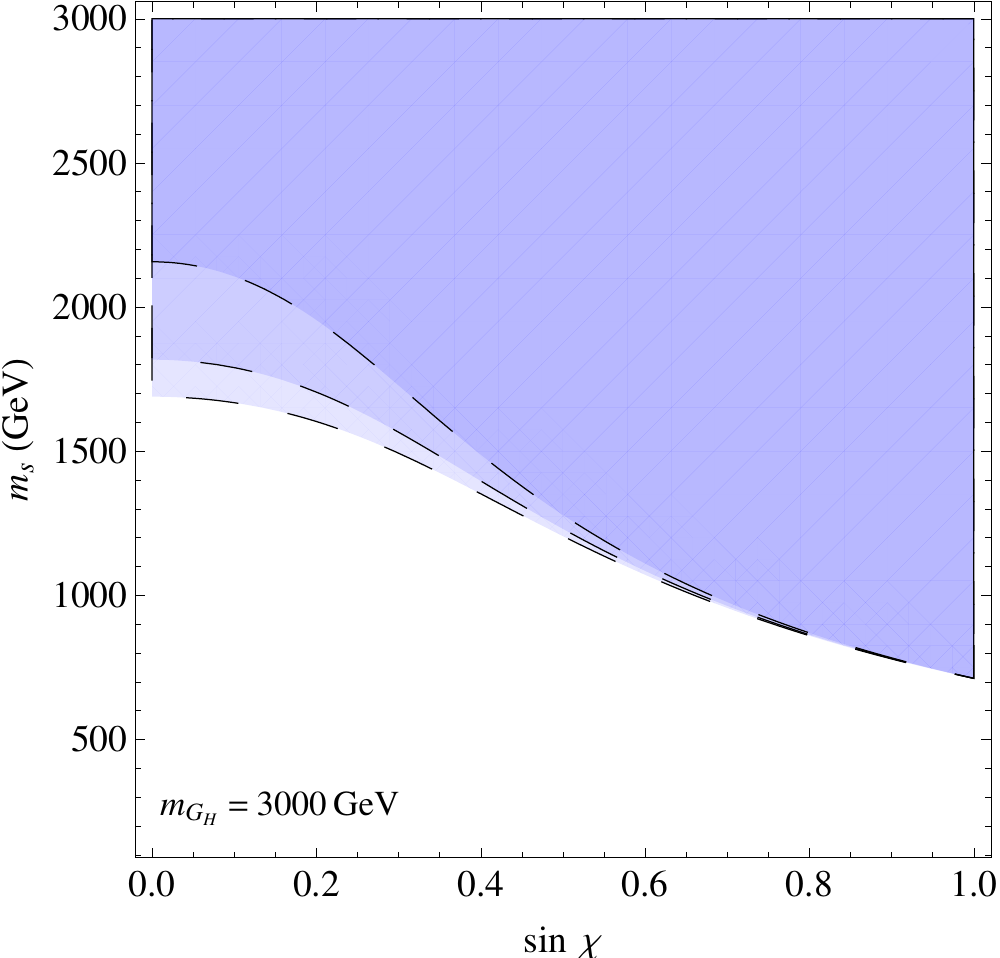}
\caption{The most stringent unitarity constraints on the $m_s-\sin \chi$ plot for $m_s\geq 150$~GeV. The top row panels represent the bounds for a singlet VEV $v_s = 500$~GeV, and for several values of the scalar color-octet mass, $m_{G_{H}}$. The panels in the middle (bottom) row correspond to $v_s = 1000\,
 (2000)$~GeV, which yield additional allowed parameter space, and also a larger $m_{G_{H}}$. In each panel, the three curves correspond (from bottom to top, $m_{\mathcal A} \in \cbrac{0,\frac{1}{2} m_{G_{H}}, m_{G_{H}}}$, and the resulting plots are superimposed, demonstrating a moderate dependence of the unitarity constraints on this parameter. A larger $m_{\mathcal A}$ slightly weakens the exclusion bounds.}
\label{a0bound}
\end{figure}

Our results are summarized in Fig.~\ref{a0bound}. The top row illustrates the exclusion plots for $v_{s} = 500$~GeV, where the panels correspond to three representative values of $m_{G_{H}}$, from light to heavy. The middle (bottom) row corresponds to $v_{s} = 1000\, (2000) $~GeV, again for three selected $m_{G_{H}}$ values. In each panel, we superimpose the constraints corresponding to three values of $m_{\mathcal A}$:  $\cbrac{0,\frac{1}{2} m_{G_{H}}, m_{G_{H}} }$, where the smallest $m_{\mathcal A}$ gives the strongest constraint. It is evident that a larger singlet VEV, accommodating heavier scalar color-octets, $G_{H}^{a}$, also provides additional allowed parameter space, whereas the sensitivity to the mass of the $\mathcal A$ pseudo-scalar is limited.

Finally, we note that the bounds obtained from the unitarity of scalar-boson scattering in the renormalizable coloron model automatically ensure that the theory remains perturbative at the energy scales of interest. In particular we have checked that, if the unitarity bounds derived in this subsetion are respected, all of the quartic couplings in the potential in eq. \eqref{pot} are (substantially) smaller than $(4\pi)^2$. Quantum loop corrections therefore remain small.

\section{Experimental Bounds on the Model}

Having explored the theoretical constraints on the renormalizable coloron model in the previous section, let us now consider the restrictions arising from the experimental data.

\subsection{Electroweak Precision Tests}
\label{sec:EWPT}

In this sub-section, we discuss how data obtained from electroweak precision tests \cite{Baak:2012kk} may be used to constrain the free parameters of the model.

As explained in section~\ref{model}, the renormalizable coloron model possesses two physical scalar degrees of freedom, $h$ and $s$, that are capable of interacting with the electroweak gauge bosons. Their couplings are, however, suppressed by factors of $\cos \chi$ and $\sin \chi$, respectively, compared with those of the SM Higgs boson, $h_{0}$ (c.f. \eqref{massbasis}). The potential effects of such a modification with respect to the ordinary SM may be explored via the oblique parameters $S$ and $T$ \cite{Peskin:1990zt,Altarelli:1990zd,Peskin:1991sw,Altarelli:1991fk}.\footnote{The contributions from the other oblique parameters are subdominant as compared with $S$ and $T$, and may be neglected.} We employ the oblique parameter expressions derived for the SM extensions containing an arbitrary number of electroweak doublet and singlet scalars\footnote{The extra spectator fermions which may be present are, as described in sec. \protect\ref{subsec:fermions}, vectorial under the electroweak interactions. Their contribution to the oblique parameters is therefore negligible.} \cite{Grimus:2008nb}; for our model, these take the form:
\begin{equation}\label{ST}
\begin{split}
S =&\, \frac{\sin^2 \chi}{24\pi} \left\{ \log R_{sh} + \hat G (m_s^2,m_Z^2) - \hat G (m_h^2,m_Z^2) \right\} \raisetag{17pt} \ , \\
T =&\, \frac{3\sin^2 \chi}{16\pi \, m_W^2 \sin^{2} \theta_W} \left\{ m_Z^2 \left[ \frac{\log R_{Zs}}{1-R_{Zs}} - \frac{\log R_{Zh}}{1-R_{Zh}} \right] - m_W^2 \left[ \frac{\log R_{Ws}}{1-R_{Ws}} - \frac{\log R_{Wh}}{1-R_{Wh}} \right] \right\} \ ,
\end{split}
\end{equation}
where, $\theta_W$ is the weak mixing angle, and we have defined
\begin{equation}\label{RGF}
\begin{split}
R_{IJ} \equiv &\, \frac{m_I^2}{m_J^2} \ , \\
\hat G(m_I^2,m_J^2) \equiv&\, -\frac{79}{3}+9R_{IJ} -2R_{IJ}^2 + \brac{12 - 4R_{IJ} + R_{IJ}^2} \hat f(R_{IJ}) + \tbrac{-10+18R_{IJ} -6R_{IJ}^2+R_{IJ}^3 - 9\frac{R_{IJ}+1}{R_{IJ}-1}}\log R_{IJ} \ , \\
\hat f(R_{IJ}) = &\,
\begin{cases}
\sqrt{R_{IJ}(R_{IJ}-4)} \, \log\left| \frac{R_{IJ}-2-\sqrt{R_{IJ}(R_{IJ}-4)}}{2} \right| &\qquad (R_{IJ}>4) \\
0 &\qquad (R_{IJ}=4)  \\
2\sqrt{R_{IJ}(4-R_{IJ})} \, \arctan\sqrt\frac{4-R_{IJ}}{R_{IJ}} &\qquad (R_{IJ}<4)
\end{cases} \ . \raisetag{30pt}
\end{split}
\end{equation}

\begin{figure}
\includegraphics[width=.49\textwidth]{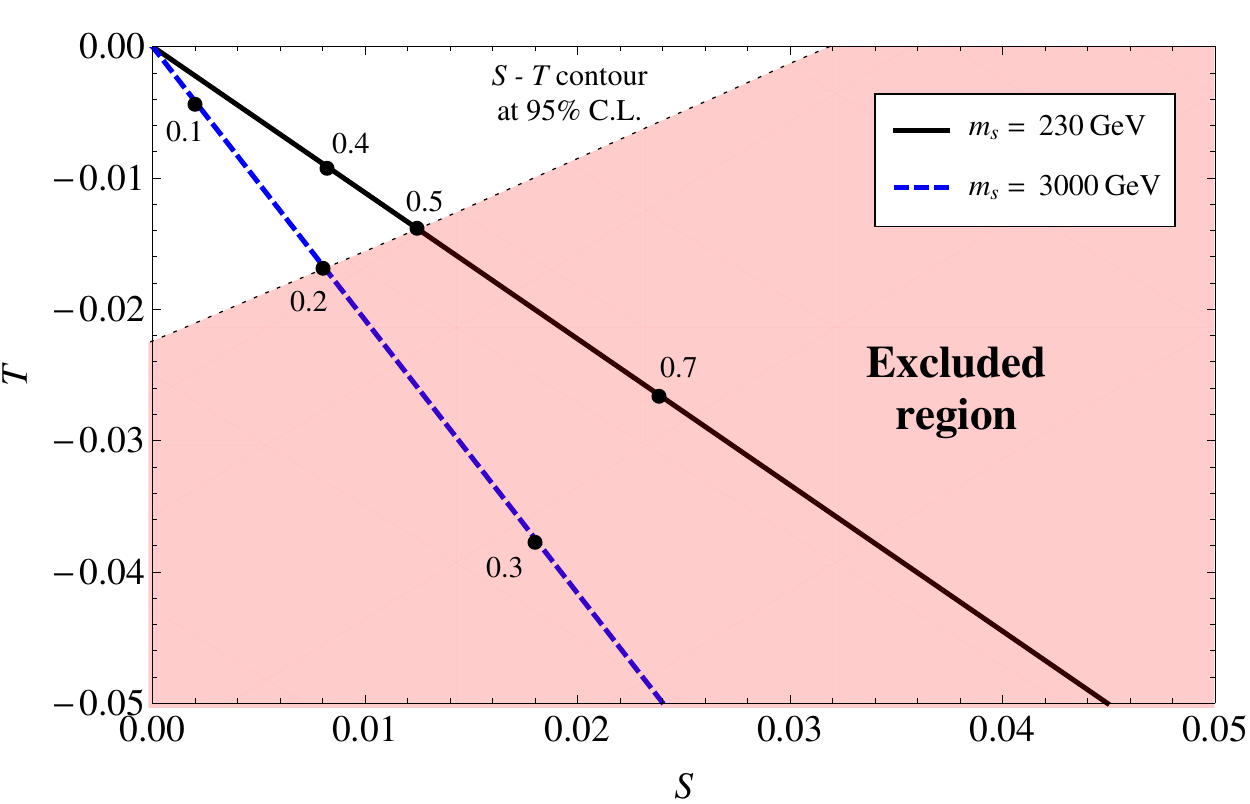}
\includegraphics[width=.49\textwidth]{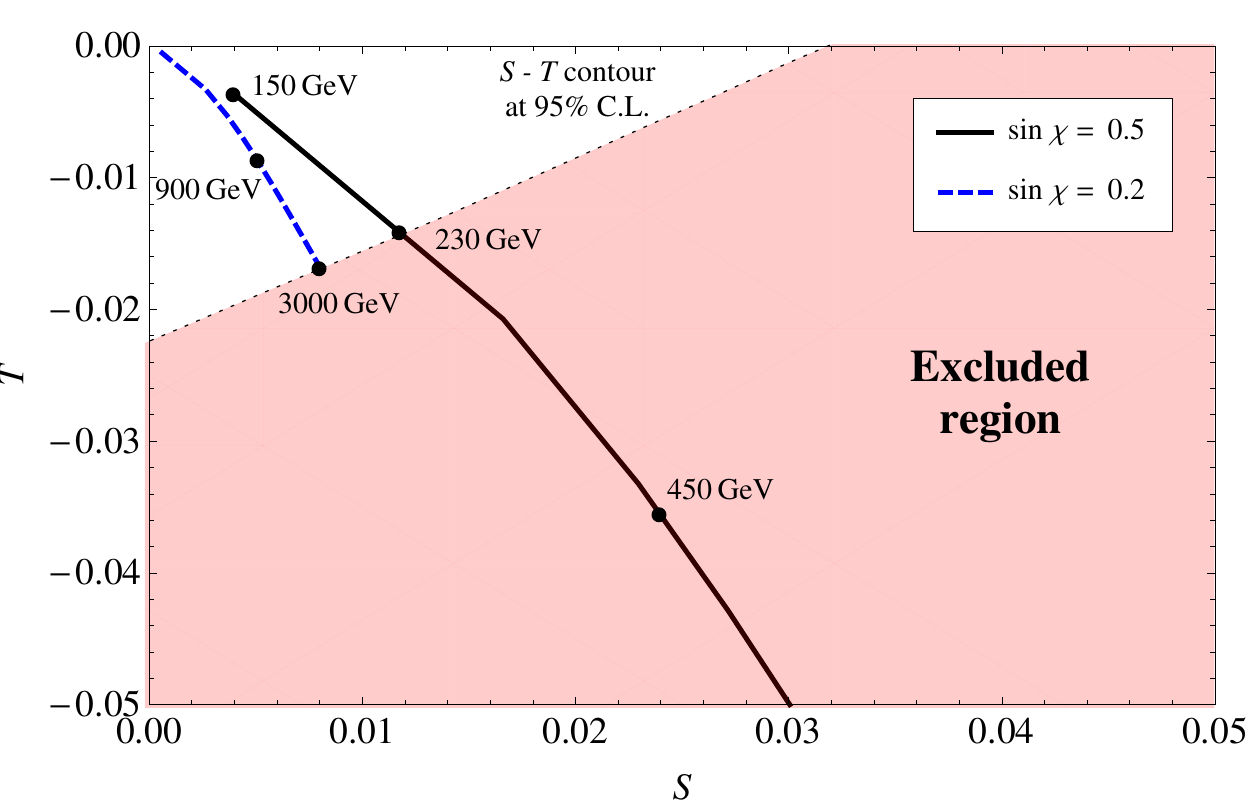}
\caption{Illustrative model predictions for the electroweak corrections $S$ and $T$, and comparison with allowed region \protect\cite{Baak:2012kk}. Note that either a larger $m_{s}$ or a larger $\sin\chi$ pushes $S$ towards more positive values and $T$ towards more negative values. \textit{Left}: Variation with scalar mixing angle for two examples of fixed $m_s$ in the $S-T$ plane with $0\leq\sin\chi\leq 1$. The 95\% C.L. contour has also been depicted, defining the allowed region. The black dots denote representative values of the mixing angle for the corresponding masses of the $s$ state, including the maximum allowed value. \textit{Right}: Variation with $m_{s}$ for two examples of fixed $\sin\chi$ with $150\leq m_s\leq 3000$~GeV. The black dots denote representative values of the $s$ scalar mass for the corresponding mixing angles, including the maximum allowed value.}
\label{STex}
\end{figure}

Following the analysis of \cite{Baak:2012kk}, 
if one sets $m_h=126$ GeV, $m_t=173$ GeV, and $M_{W,Z}$ to their observed values, then the oblique parameters \eqref{ST}
only depend on the scalar mixing angle and the mass of the $s$ scalar. 
The $S$ and $T$ values that result can be compared to the post-Higgs discovery bounds \cite{Baak:2012kk}
\begin{equation}
S=0.03 \pm 0.10~, \ \ T=0.05 \pm 0.12~, 
\end{equation}
with an $S-T$ correlation coefficient of 0.89.
As an illustration, two examples of the dependence of $S$ and $T$ on $m_s$ and $\sin\chi$ are shown in Fig.~\ref{STex}, along with the 95\% C.L. data contour.

\begin{figure}
\begin{center}
\includegraphics[width=.5\textwidth]{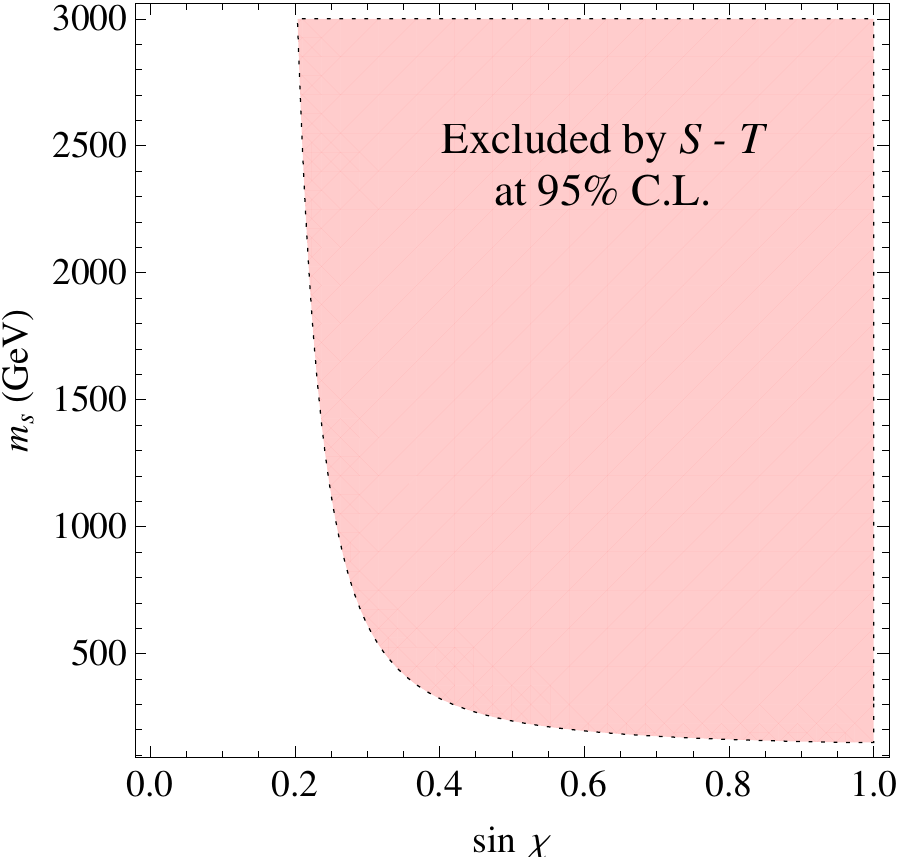}
\caption{Constraints from the electroweak precision tests in the $m_s-\sin \chi$ plane at 95\% C.L. for $m_s\geq 150$~GeV, based on the electroweak fits of \protect\cite{Baak:2012kk}. The data excludes large mixing, due to the high sensitivity of the oblique parameters to the scalar mixing angle, whereas the dependence on $m_s$ is only logarithmic for large values.}
\label{STbound}
\end{center}
\end{figure}

Let us analyze these expressions in more detail. For a heavy $s$ scalar, with a mass much larger than the electroweak scale, the function $\hat G$ in \eqref{RGF} reduces to
\begin{equation}\label{Glim}
m_s \to \infty: \qquad \hat G(m_s^2,m_Z^2) = -\frac{5}{3} +\log \frac{1}{R_{Zs}} + \mathcal{O}(R_{Zs}) \ ,
\end{equation}
which is a monotonically increasing function of $m_s^2$. Examining \eqref{ST} for a fixed mixing angle, on the one hand, reveals that in this limit, the oblique parameters retain a (mild) logarithmic dependence on $m_{s}$, albeit with the opposite signs: $S$ receives a positive contribution, whereas $T$ is pushed in the negative direction. On the other hand, keeping $m_{s}$ fixed, one notes the direct proportionality of the oblique parameters \eqref{ST} to the scalar mixing angle, once more with opposite signs. In the case of no mixing ($\sin \chi = 0$), both parameters receive no contribution from new physics, whereas in case of the maximal mixing ($\sin \chi = 1$), the oblique parameters correspond to their peak values. The oblique parameters exhibit a strong dependence on the mixing angle, while they are only mild logarithmic functions of the (large) $s$-state mass. These observations are reflected in Fig.~\ref{STbound}, which depicts the overall $S-T$ bounds at 95\% C.L. on the $m_s-\sin \chi$ plane. One notes that the electroweak precision data severely restrict the large mixing region.

\subsection{LHC Scalar Boson Results}\label{LHCsearch}

Finally, let us investigate the impact of the LHC results on the renormalizable coloron model. We begin by considering limits on the light scalar $h$, and then consider with limits on the heavy scalar $s$.

\subsubsection{Limits on the Light Scalar $h$}

 We identify the recently discovered 125~GeV Higgs-like boson with the $h$ state, resulting from a mixing among the SM Higgs boson, $h_{0}$, and the gauge-singlet scalar, $s_{0}$, as in \eqref{massbasis}. Since the light boson is an admixture of the $s_0$ and $h_0$ states, its couplings to standard model particles will be modified with respect to the expectations for a standard model Higgs boson. Furthermore, the presence of the colored scalars and vectors  and of the possible spectator fermions in the renormalizable coloron model can dramatically affect the gluon fusion cross section for this particle \cite{Kumar:2012ww}. Lastly, if the pseudo-scalar $\mathcal A$ boson is light enough, the $h \to {\mathcal A A}$ decay is open and changes the branching ratios of this particle with respect to those of the standard model Higgs.
We can parameterize the relevant $h$ couplings using an effective Lagrangian\footnote{An example of such effective Lagrangian has previously been studied for various models in \cite{Carmi:2012in} and \cite{Falkowski:2013dza}.} as\footnote{As mentioned in the beginning of section~\ref{boundsall}, the mass range $m_{G_{H}} \in \tbrac{50,125}$ is already ruled out by Tevatron searches. We therefore do not include the decay channel $h \to G_{H}^{a} G_{H}^{a}$ in our analysis.}
\begin{equation}\label{Leff}
\begin{split}
\mathcal{L}_{\text{eff}}=&\, c_V\frac{2m_W^2}{v_h}\,hW_{\mu}^{+}W^{-\mu}+c_V\frac{m_Z^2}{v_h}\,hZ_{\mu}Z^{\mu}-c_b\frac{m_b}{v_h}\,h\bar{b}b-c_{\tau}\frac{m_{\tau}}{v_h}\,h\bar{\tau}\tau-c_c\frac{m_c}{v_h}\,h\bar{c}c \\
&+c_g\frac{\alpha_s}{12\pi{v_h}}\,hG^a_{\mu\nu}G^{a\mu\nu}+c_{\gamma}\frac{\alpha}{\pi{v_h}}\,hA_{\mu\nu}A^{\mu\nu}-c_{\mathcal{A}}\,h\mathcal A \mathcal A \ .
\end{split}
\end{equation}

In the renormalizable coloron model, applying Eq.  \eqref{massbasis} yields the value of the coefficients in the first line of \eqref{Leff} 
\begin{equation}\label{SDcoup}
c_V=c_b=c_c=c_{\tau}= \cos\chi \ ,
\end{equation}
whereas, the coupling for the decay of $h$ into a pair of the pseudo-scalars, $\mathcal A$, may be determined from the scalar potential \eqref{pot} in the mass eigenstate basis to have the coupling strength
\begin{equation}\label{cinv}
c_{\mathcal{A}} = -\frac{1}{2}\frac{m_{h}^{2} + m_{\mathcal A}^{2}}{v_{s}} \sin \chi \ .
\end{equation}
The $\mathcal{A}\mathcal{A}$ decay mode becomes kinematically accessible only if $2 m_{\mathcal A} \leq m_{h} = 125$~GeV. ATLAS has set an upper limit \cite{ATLAS:2013pma} on the invisible branching fraction of a 125 GeV Higgs boson produced in association with a $Z$ boson at the SM rate; the 95\% CL upper bound established is $BR(h\to {\mathcal AA}) < 65\%$.  We find this sets no additional limits on the renormalizable coloron model if $v_s > 500$ GeV.

The leading order interaction of $h$ and a pair of gluons or photons emerges at one loop. In addition to the usual SM particles (including the top quark) with a suppressed coupling proportional to $\cos \chi$ (c.f. \eqref{SDcoup}), there are various new degrees of freedom participating in the loop-generated interaction, whose couplings are proportional to $\sin \chi$. Hence, we may parametrize the $c_{g}$ and $c_{\gamma}$ coefficients according to
\begin{equation}\label{Netcgcgamma}
c_g=\cos\chi \,\hat{c}_g^{\text{SM}}-\sin\chi \, \delta{c_g},\qquad
c_{\gamma}=\cos\chi \, \hat{c}_\gamma^{\text{SM}}-\sin\chi \, \delta{c_\gamma} \ ,
\end{equation}
where the contributions from the $t$, $b$, $c$, and $\tau$ fermions are all included in $\hat{c}_g^{\text{SM}}$ and $\hat{c}_\gamma^{\text{SM}}$ (details in Eq. \eqref{SMcoup}). The new-physics contribution to the decay of the $h$ scalar into pairs of gluons and photons are, respectively, parametrized by $\delta{c_g}$ and $\delta{c_\gamma}$ in \eqref{Netcgcgamma}. Note that the contribution of $\delta{c_g}$ ($\delta{c_\gamma}$) to $c_{g}$ ($c_{\gamma}$) is sensitive to the sign\footnote{This is the only place where such a sensitivity arises. It is evident from \eqref{conversions} that a change of sign of $\sin\chi$ corresponds to a change of sign in $\lambda_{m}$, which has no further effect on the tree-level analyses (c.f. \eqref{stab}).} of $\sin\chi$.

\begin{figure}
\begin{center}
\includegraphics[width=.8\textwidth]{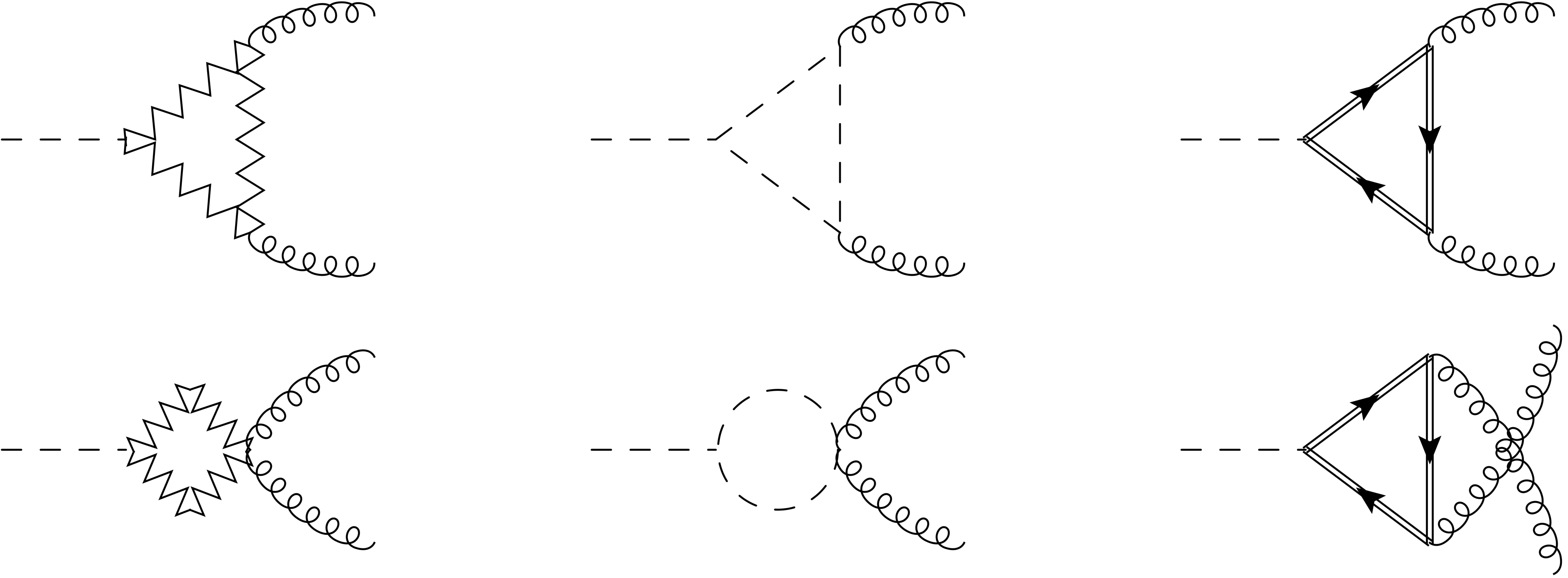}
\caption{One-loop diagrams, containing physical degrees of freedom, contributing to the new-physics contributions to  $h \to gg$. The columns represent, from left to right, the contributions of the heavy vector color-octets ($C^a_\mu$), scalar color-octets ($G_{H}^{a}$), and the spectator fermions ($Q$) in the loop.}
\label{hgg}
\end{center}
\end{figure}

\begin{figure}
\begin{center}
\includegraphics[width=.5\textwidth]{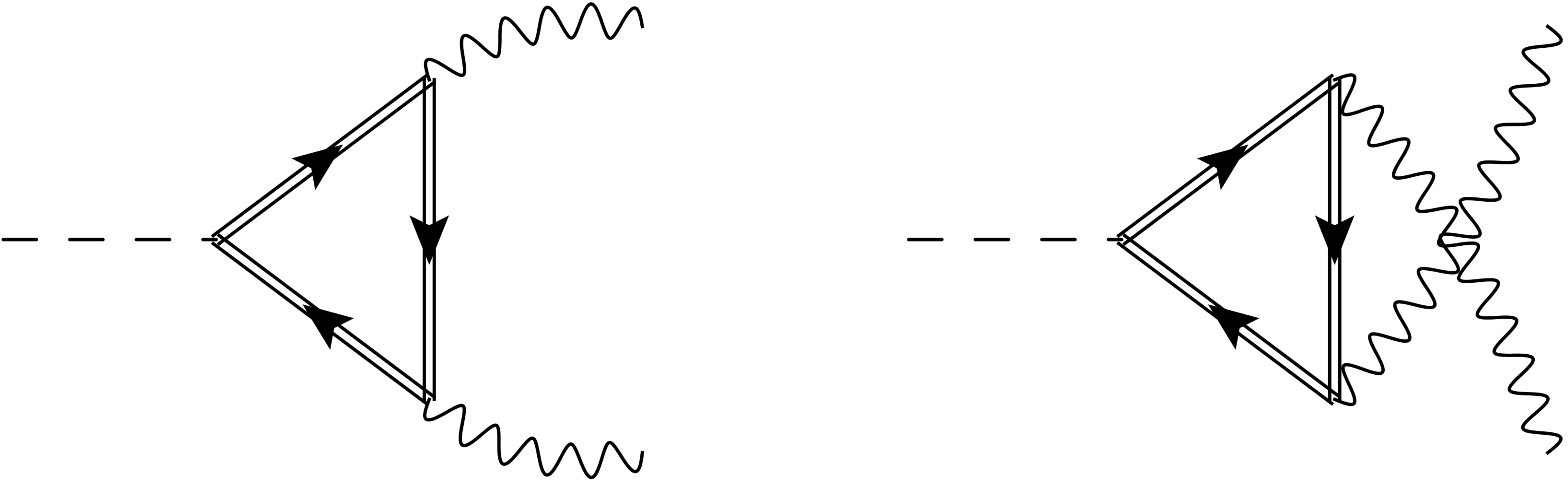}
\caption{New-physics one-loop diagrams contributing $h \to \gamma \gamma$. Only the heavy spectator fermions in the loop contribute to this process.}
\label{hAA}
\end{center}
\end{figure}

The non-SM diagrams contributing to the $h \to gg$ and $h \to \gamma \gamma$ decay processes are shown in Figs.~\ref{hgg}~and~\ref{hAA}, respectively. They represent the effects of the extra physical degrees of freedom: the vector colorons, the $G_{H}^{a}$ scalar color-octets and the spectator fermions all run in the loop for the $h \to gg$ decay process, while only the spectator fermions contribute to $h \to \gamma \gamma$. The number of spectator fermions depends on the $SU(3)_{c1} \times SU(3)_{2c}$ gauge charges chosen for the ordinary fermions. As explained in section~\ref{subsec:fermions}, three illustrative examples of the fermion charge assignments require zero, one, or three spectator quark generations -- where each spectator quark generation consists of an up- and a down-type quark. The spectator quarks have the same color and electric charges as their corresponding quark counterparts and, 
for simplicity, they are assumed to have a common mass. Therefore, all spectator flavors contribute equally to the $h \to gg$ decay, while their contributions to the $h \to \gamma \gamma$ decay channel are proportional to their corresponding electric charges.

Using the Feynman rules in Appendix~\ref{FR}, we find the couplings
\begin{align}
\delta{c_g}=&\,-3\frac{v_{h}}{v_s}\left[A_V(\tau_C)+\left(1+\frac{m_h^2-\frac{2}{3}m_\mathcal A^2}{2m_{G_H}^2}\right)A_S(\tau_{G_H})-\frac{2 N_Q}{3}A_F(\tau_Q)\right] \ , \label{delcg}\\
\delta{c_\gamma}=&\, \frac{5N_Q}{18}\frac{v_{h}}{v_s}A_F(\tau_Q) \label{delcA} \ ,
\end{align}
where $N_Q$ is the number of spectator generations, $\tau_i\equiv \dfrac{m^2_h}{4m_i^2}$, and the subscript $C$ ($Q$) represents the coloron (spectator). In addition, we have defined the vector, fermion, and scalar form factors, respectively, as
\begin{equation}\label{formfact}
\begin{split}
A_V(\tau)\equiv&\,\frac{1}{8\tau^2}\tbrac{3\tau+2\tau^2- 3(1-2\tau)f(\tau)} \ , \quad
A_F(\tau)\equiv \frac{3}{2\tau^2}\tbrac{\tau- (1-\tau)f(\tau)} \ , \quad
A_S(\tau)\equiv\frac{1}{8\tau^2}\tbrac{\tau-f(\tau)} \ , \\
f(\tau) \equiv&\,
\begin{cases}
\arcsin^{2}\sqrt{\tau} \qquad \qquad \qquad \qquad \; \,\tau\leq1\\
-\frac{1}{4}\tbrac{\log \frac{1+\sqrt{1-\tau^{-1}}}{1-\sqrt{1-\tau^{-1}}}-i\pi}^2 \qquad \tau>1
\end{cases} \ .
\end{split}
\end{equation}
For completeness, we also note the SM contributions
\begin{equation}\label{SMcoup}
\begin{split}
\hat{c}_g^{\text{SM}}\equiv&\, A_F(\tau_t)+A_F(\tau_b)+A_F(\tau_c) \ ,\\
\hat{c}_\gamma^{\text{SM}}\equiv&\,-A_V(\tau_W)+\frac{1}{18}A_F(\tau_b)+\frac{2}{9}\tbrac{A_F(\tau_t)+A_F(\tau_c)}+\frac{1}{6}A_F(\tau_{\tau}) \ ,
\end{split}
\end{equation}
which include the contributions from the $t$, $b$, $c$, and $\tau$ fermions. The vector, fermion, and scalar form factors \eqref{formfact} quickly converge to their asymptotic values, once their corresponding massive particles are heavier than $ m_{h}=125$~GeV,
\begin{equation}\label{asymp}
M_{C}, M_{Q}, m_{G_{H}} \gtrsim m_{h} \qquad \Longrightarrow \qquad A_V\to\frac{7}{8},\quad A_F\to1,\quad A_S\to-\frac{1}{24} \ ,
\end{equation}
rendering \eqref{delcg} insensitive to the precise values of these parameters in the heavy mass limit.

Having determined all of the coupling coefficients in the Lagrangian \eqref{Leff}, we may now incorporate the available LHC experimental data in a global-fitting analysis. The Lagrangian coefficients are functions of all of the model's free parameters \eqref{freepar}, except  $m_{s}$.
As mentioned, the sensitivity to  the parameters $M_{C}$, $M_{Q}$, and $m_{G_{H}}$ is negligible for the region of interest, due to the asymptotic behavior \eqref{asymp}.
Hence, to construct an $m_s-\sin \chi$ exclusion plot, we need to find the best-fit value of the scalar mixing angle, $\sin\chi$, using the remaining relevant parameters as input. 

\begin{table}[t]
\begin{center}
\begin{tabular}{|c|c|c|c|c||c|c|}
\hline
 DATA & $\mu _{\textrm{ggF+ttH}}$ & $\mu _{\textrm{VBF+VH}}$ & $\rho$  & $\mu _{\textrm{comb}}$ & $\mu^{\textrm{exp}} _{\textrm{comb}}$ & Ref\\
\hline
 CMS $\gamma \gamma $ & $0.468\pm0.396$ & $1.69\pm0.868$ & -0.485 & $0.79\pm0.27$ & $0.77\pm0.27$ & \\
 CMS ZZ & $0.958\pm0.432$ & $1.29\pm2.13$ & -0.671 & $1.0\pm0.28$ & $0.92\pm0.28$ & \\
 CMS WW & $0.763\pm0.233$ & $0.331\pm0.692$ & -0.246 & $0.70\pm0.20$ & $0.68\pm0.20$ & \cite{CMS:yva}\\
 CMS $\tau \tau $ & $0.684\pm0.793$ & $1.61\pm0.825$ & -0.469 & $1.13\pm0.42$ & $1.10\pm0.41$ & \\
 CMS bb & $0.477\pm2.62$ & $1.25\pm0.651$ & 0.0051 & $1.21\pm0.63$ & $1.15\pm0.62$ & \\
\hline
 ATLAS $\gamma \gamma $ & $1.62\pm0.411$ & $1.92\pm0.819$ & -0.275 & $1.69\pm0.32$ & $1.6\pm0.3$ & \\
 ATLAS ZZ & $1.49\pm0.517$ & $1.84\pm1.91$ & -0.491 & $1.54\pm0.39$ & $1.5\pm0.4$ &  \\
 ATLAS WW & $0.793\pm0.344$ & $1.69\pm0.758$ & -0.186 & $0.98\pm0.29$ & $1.0\pm0.3$ & \cite{ATLAS:2013sla} \\
 ATLAS $\tau \tau $ & $2.28\pm1.48$ & $-0.189\pm1.04$ & -0.435 & $0.75\pm0.65$ & $0.8\pm0.7$ & \\
\cline{2-5}
 ATLAS bb  & \multicolumn{4}{c||}{None} & $-0.4\pm1.0$ & \\
\hline
 Tevatron bb & \multicolumn{4}{c||}{None} & $1.56\pm0.72$ & \cite{Tevatron:2013}\\
\hline
\end{tabular}
\caption{\label{tab:LHCData} LHC and Tevatron data on the properties of the newly observed scalar boson, expressed as measured cross section times branching ratio relative to the standard model Higgs boson. The values for the strengths of the separate gluon fusion ($\mu _{\textrm{ggF+ttH}}$) and vector-boson fusion ($\mu _{\textrm{VBF+VH}}$) production mechanisms, and their correlations ($\rho$) are read from the plots presented in the references given. We have checked that the total signal strength that results from our values ($\mu _{\textrm{comb}}$) agrees well with the numerical result quoted  ($\mu^{\textrm{exp}} _{\textrm{comb}}$). For both ATLAS bb and Tevatron bb, $\mu^{\textrm{exp}}_{\textrm{comb}}$ denotes VH production.}
\end{center}
\end{table}

The data we have used in our fit to the properties of the $h$ is displayed in Table \ref{tab:LHCData}, and is derived from \cite{CMS:yva,ATLAS:2013sla,Tevatron:2013}.
In order to make our fit to be more informative, we have read the strengths of the separate gluon fusion ($\mu _{\textrm{ggF+ttH}}$) and vector-boson fusion ($\mu _{\textrm{VBF+VH}}$) production mechanisms, and their corresponding correlation coefficient ($\rho$) from the plots presented in \cite{CMS:yva,ATLAS:2013sla}. As a cross check, we have verified that the total signal strength that results from our values ($\mu _{\textrm{comb}}$) agrees well with the numerical result quoted  ($\mu^{\textrm{exp}} _{\textrm{comb}}$).

\begin{figure}[!h]
    \includegraphics[width=16cm]{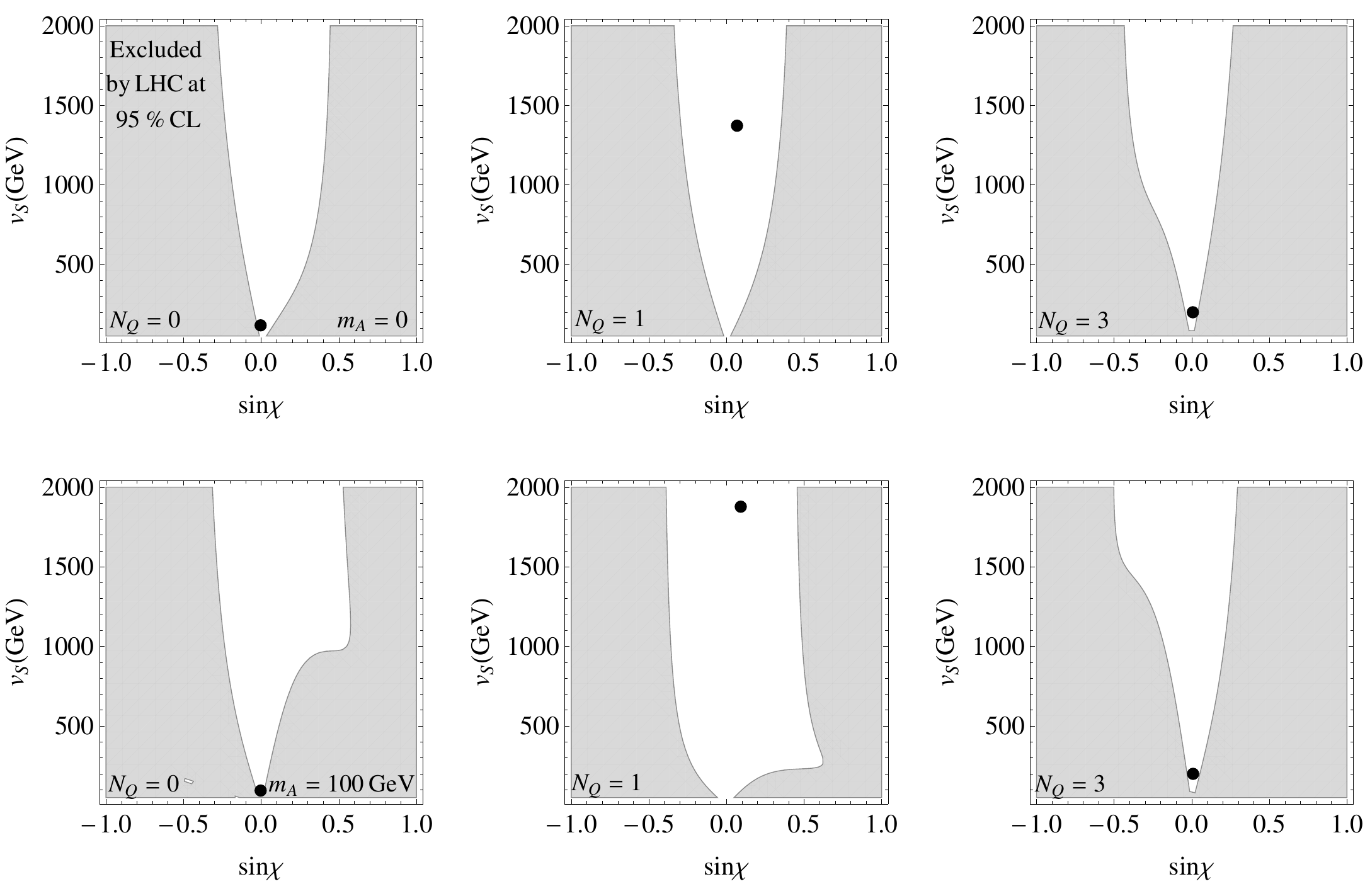}
\caption{\label{fig:LHC95CLExcmA0mA100} The constraints on the $v_s$ -- $\sin\chi$ plane derived from the properties of the 125 GeV scalar observed by the LHC; the grey region is excluded at 95\% CL. We choose $m_{\mathcal A}=0$, which allows for the invisible decay $h\to {\mathcal AA}$, in the first row. and $m_{\mathcal A}=100$ GeV (which forbids that invisible decay) in the second row. The three columns are for zero, one, and three generations of spectator quarks, reading from left to right. Note that the sign of $\sin\chi$ matters.}
\end{figure}

Fig.~\ref{fig:LHC95CLExcmA0mA100} displays the
best fit values for two choices of the pseudo-scalar mass $m_{\mathcal A}$. In the first row, we display the values for $m_{\mathcal A}=0$, which is representative of our results when the invisible decay $h \to {\mathcal AA}$ is open.
In this case, the presence of the invisible decay mode uniformly decreases all of the visible signal strengths relative to standard model expectations. This suppression disfavors the additional suppression of these decays which would occur through substantial mixing ({\it i.e.} small $\cos\chi$), resulting in the best fit region's favoring small $\sin\chi$. 

In the lower row, however, no invisible decay mode is present. Depending on the sign of $\sin\chi$, the additional contributions from the spectator fermions can interfere constructively or destructively with the standard model contributions, both in the gluon fusion production mechanism and in the diphoton decay amplitudes. In this case, there is a much larger model-dependence on the bounds depending on the spectator content of the theory. 

Finally, given the curent uncertainties in the observations, the $\chi^2$ minimum is relatively shallow
and the ``best-fit" values in Fig. \ref{fig:LHC95CLExcmA0mA100} change significantly across the six diagrams.  Taken together, however, these diagrams are a fair representation of the region in $\sin\chi$ and $v_s$ which is allowed by current LHC data.

\subsubsection{LHC Limits on the Heavy Scalar $s$}

Since the heavy $s$ is an admixture of the gauge-eigenstate $s_0$ with the $h_0$ fields, it would appear in searches for a heavy ``Higgs" boson in the $W^+W^-$ and $ZZ$ channels at the LHC \cite{CMSHeavyH,ATLASHeavyH}. Therefore, as we will see, the LHC is potentially sensitive to an $s$-boson as heavy as 1 TeV. On the other hand, collider searches require the coloron mass, $M_{C}$, to be at least in the TeV region \cite{Simmons:1996fz,Bertram:1998wf,ATLAS:2012pu,ATLAS:2012qjz,Chatrchyan:2013qha,CMS:kxa} and, hence, we can neglect the decays of $s$ to pairs of colorons. In addition, in what follows, we will also assume that any spectator quarks $Q$ are too heavy to be produced in $s$-boson decay.  The couplings of the $Q$ to $s$ arise from the terms in (\ref{Lferm}), and are therefore proportional to $M_Q$; in consequence, the contribution of the spectator fermions to $s$ production from gluon
fusion does not ``decouple" as $M_Q$ becomes large. By assuming that $M_Q$ is large, we maximize the
$WW$ and $ZZ$ branching ratios of $s$ without significantly suppressing the production cross section -- and hence 
the limits we discuss below reflect the {\it maximum} LHC sensitivity  for this particle.

In this regime, by analogy with (\ref{Leff}), we can write an effective Lagrangian for $s$ phenomenology at the LHC,
\begin{eqnarray}
\label{heavyLeff}
\mathcal{L}_{s,\text{eff}} &=&
   c^s_V \frac{2m_W^2}{v_h}\,s\,W_{\mu}^{+}W^{-\mu}
  +c^s_V \frac{m_Z^2}{v_h}\,s\,Z_{\mu}Z^{\mu}
  -c^s_t \frac{m_t}{v_h}\,s\,\bar{t}t
  -c^s_b \frac{m_b}{v_h}\,s\,\bar{b}b
  -c^s_\tau \frac{m_\tau}{v_h}\,s\,\bar{\tau}\tau
  -c^s_c \frac{m_c}{v_h}\,s\,\bar{c}c
\nonumber\\
&&
  +c^s_g\frac{\alpha_s}{12\pi{v_h}}\,s\,G^a_{\mu\nu}G^{a\mu\nu}
  +c^s_\gamma\frac{\alpha}{\pi{v_h}}\,s\,A_{\mu\nu}A^{\mu\nu}
  +c^s_h \,s\,h h
  +c^s_\mathcal{A} \,s\,\mathcal{A}\mathcal{A}
  +c^s_{G_H} \,s\,G_H^a G_H^a~,
\end{eqnarray}
where we include the additional possibilities that the heavy scalar may decay to pairs of the light scalar $h$, top-quarks, and the additional colored scalars. The tree level couplings of SM particles to $s$ are suppressed by the mixing angle,
and hence
\begin{eqnarray}
c^s_V=c^s_t=c^s_b=c^s_\tau=c^s_c=\sin\chi~.
\end{eqnarray}

Using the Feynman rules in Appendix~\ref{FR}, we find that the effective coupling of the $s$ to gluons is,
\begin{eqnarray}\label{eq:cgs}
c^s_g&=&\sin\chi \,\hat{c}_g^{\text{SM}}+\cos\chi \, \hat{c}^s_g,\nonumber\\
\hat{c}_g^{\text{SM}}&\equiv&\, A_F(\tau^s_t)+A_F(\tau^s_b)+A_F(\tau^s_c)\nonumber\\
\hat{c}^s_g&\equiv&-3\frac{v_{h}}{v_s}\left[A_V(\tau^s_C)+\left(1+\frac{m_s^2-\frac{2}{3}m_\mathcal A^2}{2m_{G_H}^2}\right)A_S(\tau^s_{G_H})-\frac{2 N_Q}{3}A_F(\tau^s_Q)\right]
\end{eqnarray}
where $\tau^s_i\equiv m_s^2/4m_i^2$ and $N_Q$ is the number of spectator quark generations. Note that here, unlike in the case of the $h$ considered above, the contributions of the extra colored scalar, vector, and fermionic states are potentially quite important since $\cos \chi$  can be
sizable. Note also that there is a possible cancellation between the fermion and coloron contributions (the scalar term can be of either sign) -- a feature that will be important in our discussion below. On the other hand, the contribution of the standard model particles to gluon fusion production of $s$ is quite small.

We may derive expressions for the other relevant scalar couplings from the potential:
\begin{eqnarray}
c_h^s
&=&-\frac{\sin\chi\cos\chi}{2v_hv_s}\left[v_h\left(\frac{m_{\mathcal{A}}^2}{3}+2m_h^2+m_s^2\right)\sin\chi+v_s(2m_h^2+m_s^2)\cos\chi\right]\nonumber\\
c_\mathcal{A}^s
&=&-\frac{m_\mathcal{A}^2+m_s^2}{2v_s}\cos\chi\nonumber\\
c_{G_H}^s&=&-\frac{m_s^2+2m_{G_H}^2-\frac{2}{3}m_\mathcal{A}^2}{2v_s}\cos\chi~.
\end{eqnarray}
The associated decay width for $i=h, \mathcal{A}, G_H^a$ is
\begin{eqnarray}
\Gamma(s\to i i)=\frac{(c_i^s)^2}{8\pi m_s}\sqrt{1-\frac{4m_i^2}{m_s^2}},\quad
\end{eqnarray}
with $c_i^s=c_h^s, c_\mathcal{A}^s, c_{G_H}^s$ respectively. As we will see, when kinematically allowed, the decay of the $s$ to $\mathcal{A}, G_H$ is dominant in the small mixing region.

Results from the LHC \cite{CMSHeavyH,ATLASHeavyH} place constraints on the existence of the $s$ by limiting the signal strength in the $W^+W^-$ and  $ZZ$ channels from a ``heavy Higgs" boson. In the narrow width approximation, these measurements place upper bounds on the ratio
of production cross section times branching ratio of the $s$ boson relative to a standard model Higgs of the same mass
\begin{equation}
\mu_{mode}(ii\to s\to kk)=\frac{\sigma(pp\to ii \to s)}{\sigma(pp\to ii \to H^{\textrm{SM}})}\frac{\textrm{BR}(s \to kk)}{\textrm{BR}(H^{\textrm{SM}} \to kk)}
\end{equation}
where $mode$ is either gluon fusion (ggF) or vector boson fusion (VBF), $ii$ denotes the particles fusing to produce the scalar ($gg$ in the ggF  mode or $VV$ in the VBF mode) and $kk$ denotes the decay products.
The ratio of production cross sections for ggF and VBF are
\begin{eqnarray}
\frac{\sigma(pp\to gg \to s)}{\sigma(pp\to gg \to H^{\textrm{SM}})}=\left|\frac{\hat{c}_g^s}{\hat{c}_g^{\textrm{SM}}}\right|^2\equiv\kappa^s_{gg},\quad\quad
\frac{\sigma(pp\to VV \to s)}{\sigma(pp\to VV \to H^{\textrm{SM}})}=\sin^2\chi~.
\end{eqnarray}
As described above, small mixing angle $\sin\chi$  is preferred in order to identify the 125 GeV signal as our $h$ particle. Therefore, since the heavy particles contribute significantly to $\hat{c}_g^s$,
gluon fusion production of $s$ dominates over vector boson production.
We therefore focus our attention on,
\begin{eqnarray}\label{ggFsVV}
\mu_{ggF}(\textrm{gg}\to s\to VV)=\left|\frac{\hat{c}_g^s}{\hat{c}_g^{\textrm{SM}}}\right|^2\frac{\textrm{BR}(s\to VV)}{\textrm{BR}(H\to VV)_{\textrm{SM}}}
=\left|\frac{\hat{c}_g^s}{\hat{c}_g^{\textrm{SM}}}\right|^2 \sin^2\chi  \left[ \frac{\Gamma_s^{TOT}(m_s)}{\Gamma_H(m_s)}\right]^{-1} ~.
\end{eqnarray}
where, including the gluon, vector boson, light fermion, and scalar decays
of the $s$, the ratio of total widths is
\begin{eqnarray}\label{Ctots}
\frac{\Gamma_s^{TOT}(m_s)}{\Gamma_H(m_s)} =\left[\left|\frac{\hat{c}^s_g}{\hat{c}_g^{\textrm{SM}}}\right|^2\textrm{BR}^{\textrm{SM}}_{gg}
+\sin^2\chi(\textrm{BR}^{\textrm{SM}}_{VV}+\textrm{BR}^{\textrm{SM}}_{\bar{f}f})+\frac{\Gamma(s\to hh)+\Gamma(s\to\mathcal{A}\mathcal{A})+8\Gamma(s\to G_HG_H)}{\Gamma_H(m_s)}\right]~,
\end{eqnarray}
and where $VV=WW, ZZ$ and $\bar{f}f=\bar{t}t, \bar{b}b, \bar{c}c, \bar{\tau}\tau$, and where $\Gamma_{H}(m_s)$ is the SM Higgs width with $m_H=m_s$.

\begin{figure}[!h]
  \centering%
    \includegraphics[width=.47\textwidth]{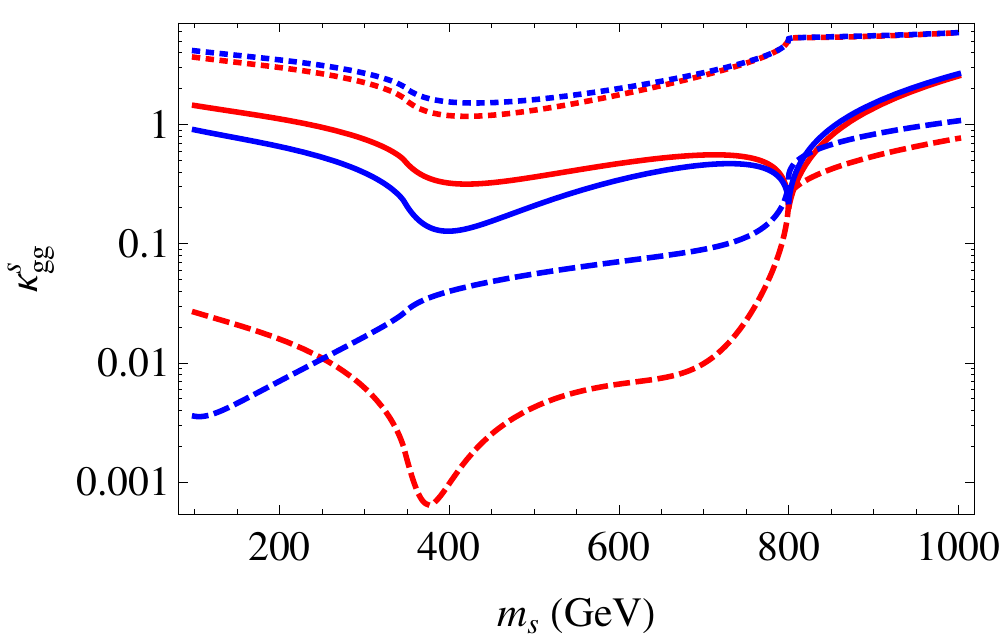}
    \includegraphics[width=.47\textwidth]{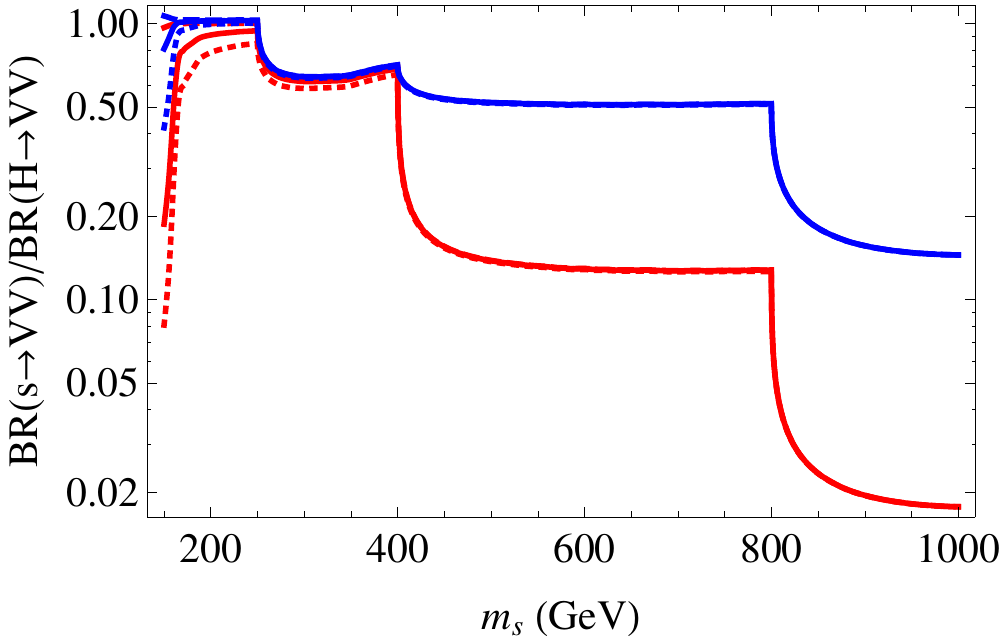}
\caption{\label{fig:ksggBRratio} For the parameter set eq.(\ref{eq:parameterset}). Left: $\kappa_{gg}^s$, the ratio of the gluon-fusion production cross section of $s$ relative to that for a standard model Higgs of mass $m_s$, as a function of $m_s$ for a given set of $\sin\chi, N_Q$. The red and blue lines are for $\sin\chi=0.1, 0.3$, and thick, dashed and dotted lines for $N_Q=0, 1, 3$, respectively. Right: $\textrm{BR}(s\to VV)/\textrm{BR}(H\to VV)$ as a function of $m_s$ for different $\sin\chi$, where the red and blue lines are for $\sin\chi=0.1, 0.3$, respectively.}
\end{figure}

In figure \ref{fig:ksggBRratio} we plot the production cross section and diboson branching ratio of the $s$-boson, relative to those for a standard model Higgs boson of the same mass, for an illustrative choice of parameters
\begin{eqnarray}\label{eq:parameterset}
m_\mathcal{A}=200\,\textrm{GeV},\,m_{G_H}=400\,\textrm{GeV},\, M_C=2\,\textrm{TeV},\, M_Q=1\,\textrm{TeV}, \,  v_s=500\,\textrm{GeV}~,
\end{eqnarray}
for $\sin\chi=0.1$ and 0.3, and for $N_Q=0,\, 1$ and 3. Considering production (shown on the left hand side of figure \ref{fig:ksggBRratio}), we see that the gluon fusion production cross section of the $s$-boson is sizable both for $N_Q=0$, dominated by the contribution from colored scalar- and vector-boson loops, and for $N_Q=3$, dominated by the contribution from spectator quarks. In the case $N_Q=1$, however, there is substantial cancellation between these contributions -- and the cross section is substantially {\it smaller}. There are also inflection points of these cross sections at the thresholds $2m_t$ and $2m_{G_H}$, where the corresponding gluon fusion amplitude acquires an imaginary part, and this accounts for the increase in the $N_Q=1$ cross section for masses above 800 GeV. The dependence of the production cross section on $\sin\chi$, on the other hand, is relatively mild, except in the case that $N_Q=1$, in which the amount of cancellation is sensitive to the top-quark contribution (which is $\sin\chi$-dependent).

The ratio of branching ratios, as shown on the right hand side of figure \ref{fig:ksggBRratio}, behaves quite differently. Here the dependence on $N_Q$ is almost entirely absent, since we are in a regime where the $s$ boson cannot decay into pairs of spectator quarks. As we saw, the partial width of the $s$-boson to decay to $WW$ or $ZZ$ is proportional to $\sin^2\chi$ and, therefore, when other decay channels open the branching ratio of the $s$ to dibosons is much smaller for $\sin\chi=0.1$ than it is for $\sin\chi=0.3$. The diboson branching ratio of the $s$ drops as the mass $m_s$ crosses the thresholds for $hh$, $t\bar{t}$, ${\cal AA}$, and $G_H G_H$.
Since the coupling of the $s$-boson to the top-quark is also proportional to $\sin\chi$, the diboson branching ratio of the $s$ is  independent  of $\sin\chi$ below $2m_{\cal A}$, but then drops precipitously -- especially for smaller values of $\sin\chi$.

\begin{figure}[!h]
  \centering%
    \includegraphics[width=.329\textwidth]{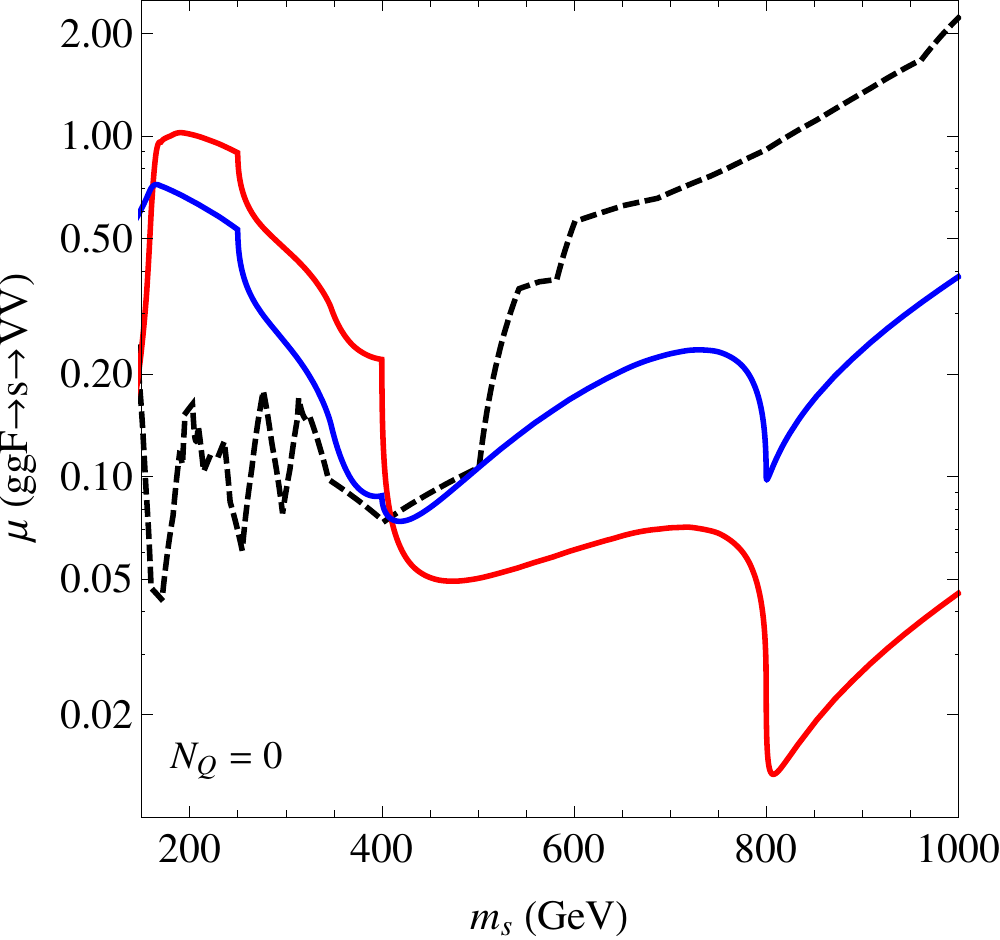}
    \includegraphics[width=.329\textwidth]{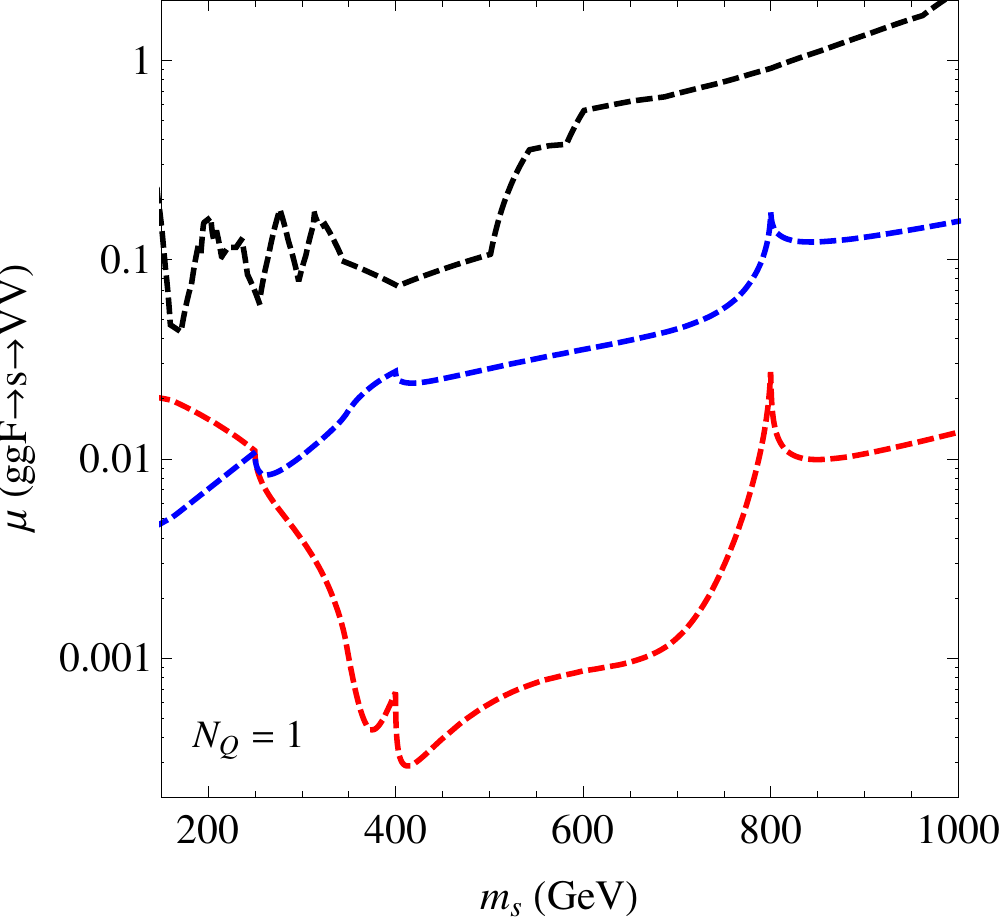}
    \includegraphics[width=.329\textwidth]{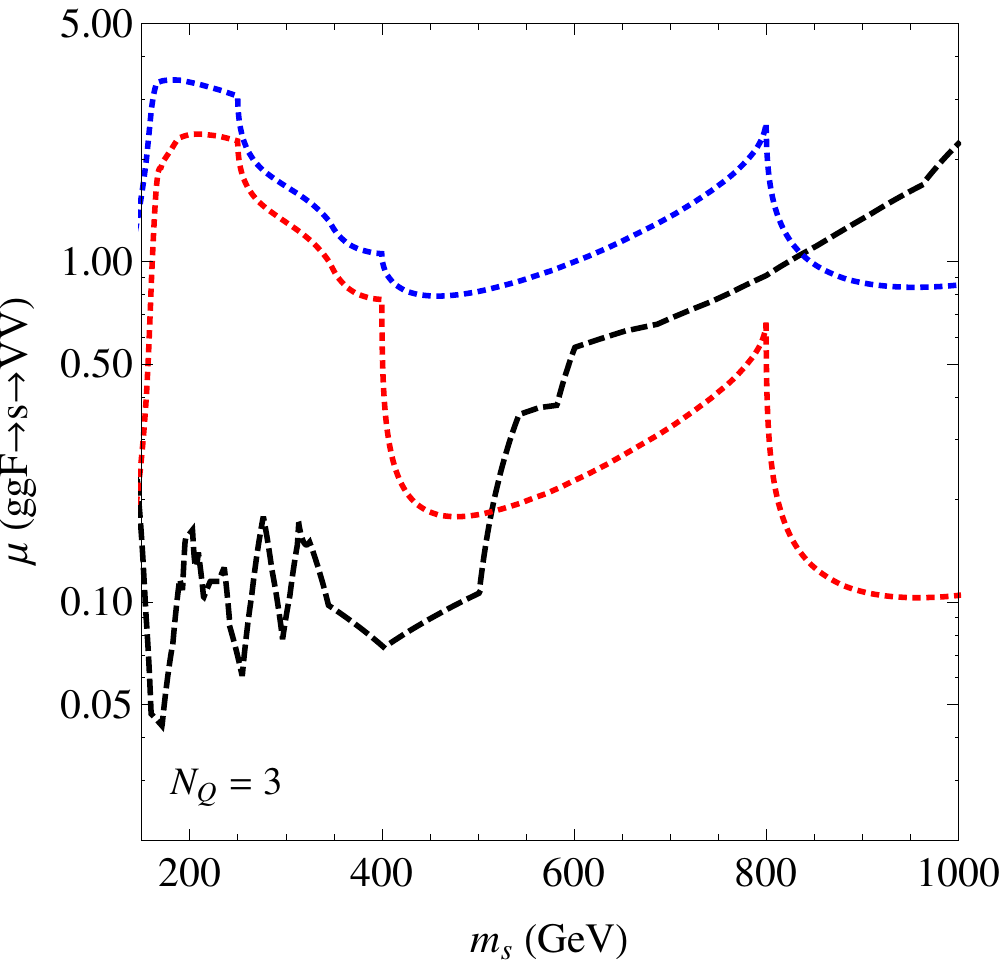}
\caption{\label{fig:muggFsVV} The black dashed line denotes the strongest exclusion at 95\% C.L. from $H\to ZZ$ and $H\to WW$ channels at CMS\cite{CMSHeavyH} and ATLAS\cite{ATLASHeavyH} respectively.  Using the model parameter values of eq.(\ref{eq:parameterset}),  the red and blue lines in each panel are for $\sin\chi=0.1, 0.3$; and the panels from left to right are for $N_Q=0, 1, 3$ respectively.}
\end{figure}

Finally, we overlay the production ratios $\mu$ (i.e., cross section times branching fraction for the $s$, compared with that of a standard model Higgs of mass $m_s$) with experimental results in Fig.\ref{fig:muggFsVV}. The region above the black dashed lines is excluded at $95\% $C.L. by $H\to ZZ$ (up to 1TeV) and $H\to WW$ (up to 600 GeV) search at CMS \cite{CMSHeavyH} and ATLAS \cite{ATLASHeavyH} respectively. Using the benchmark parameters in eq.(\ref{eq:parameterset}), the red and blue lines are for $\sin\chi=0.1, 0.3$, and the panels from left to right are for $N_Q=0, 1, 3$ respectively. Here, we can see the interplay of production cross section and branching ratio. 
For $N_Q=0$, current LHC searches exclude (for the assumed parameters)
$m_s \stackrel{<}{\sim} 400$ GeV. For $N_Q=3$, the production cross section is large enough that $m_s\stackrel{<}{\sim} 800$ GeV is excluded for $\sin\chi=0.3$ whereas, due to the
lower diboson branching ratio, only $m_s\stackrel{<}{\sim}550$ GeV is excluded for $\sin\chi=0.1$. 
Finally, for $N_Q=1$, the small cross section (due to the near-cancellation of the coloron and fermion contributions to ggF production of $s$)  leaves us with {\it no bound} on $m_s$ from current LHC data.

As this brief discussion illustrates, therefore, the LHC is potentially sensitive to the heavy singlet scalar $s$-boson in the diboson searches for a heavy ``Higgs" boson. The signal rate for the $s$-boson, however, is highly dependent on the parameters of the model -- on the mixing angle $\sin\chi$, on the spectrum of additional scalar particles, and especially on the number of spectator quark generations present. A more complete investigation of this signal is underway \cite{new}.

\section{Summary}

\begin{figure}
\includegraphics[width=.329\textwidth]{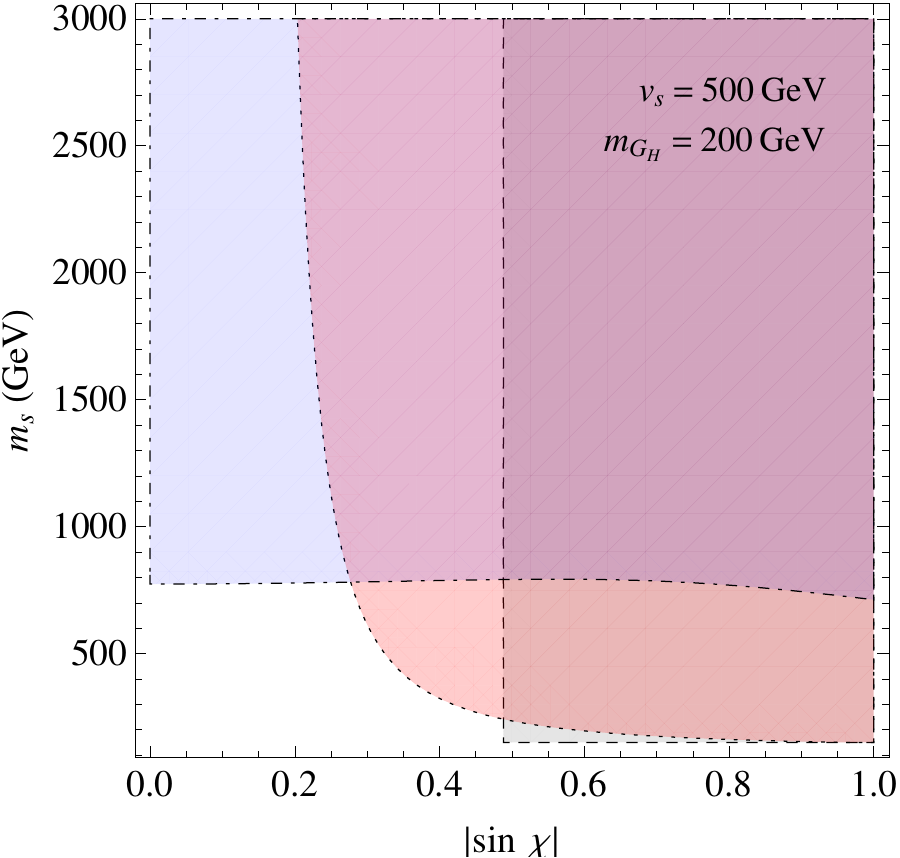}
\includegraphics[width=.329\textwidth]{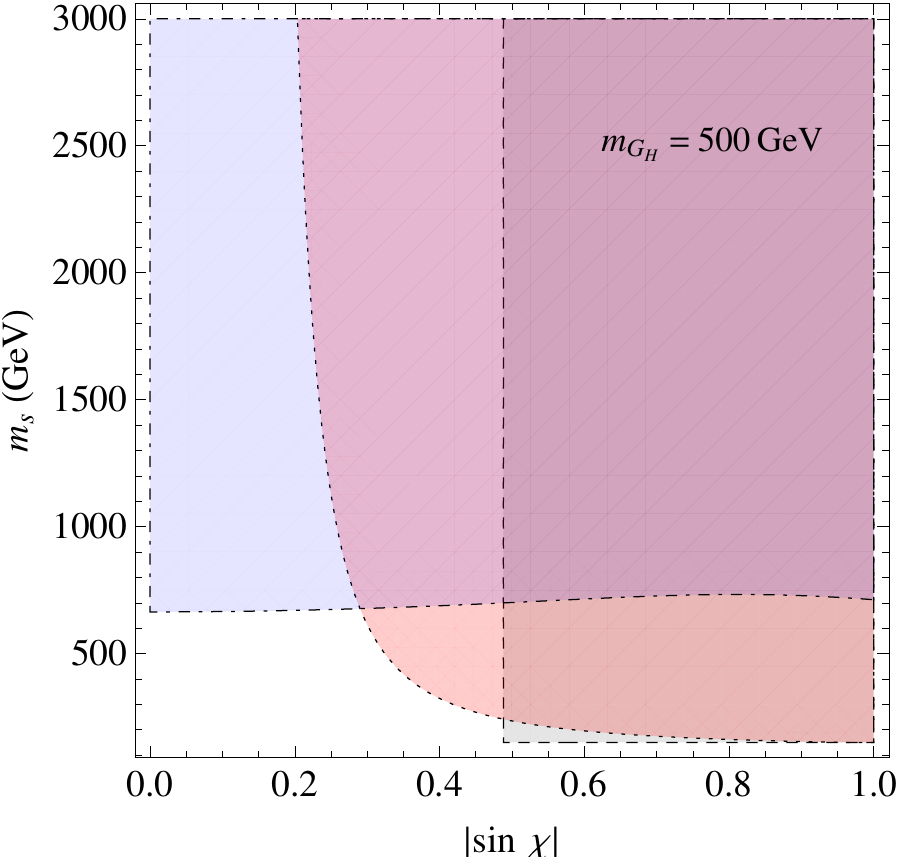}
\includegraphics[width=.329\textwidth]{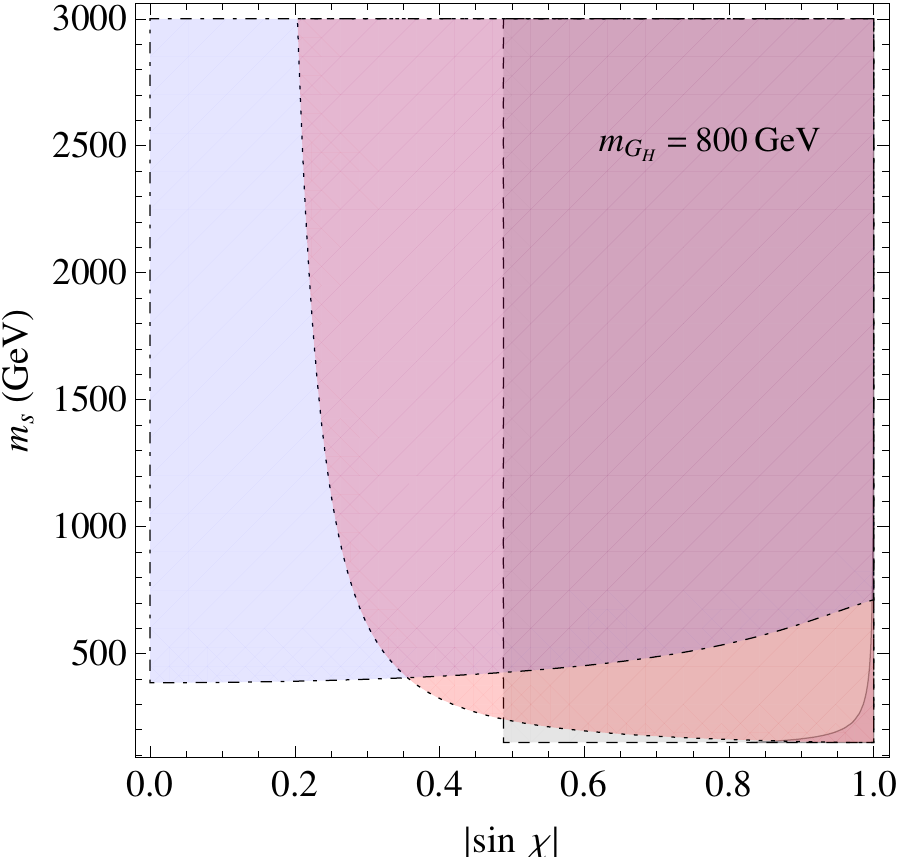}
\includegraphics[width=.329\textwidth]{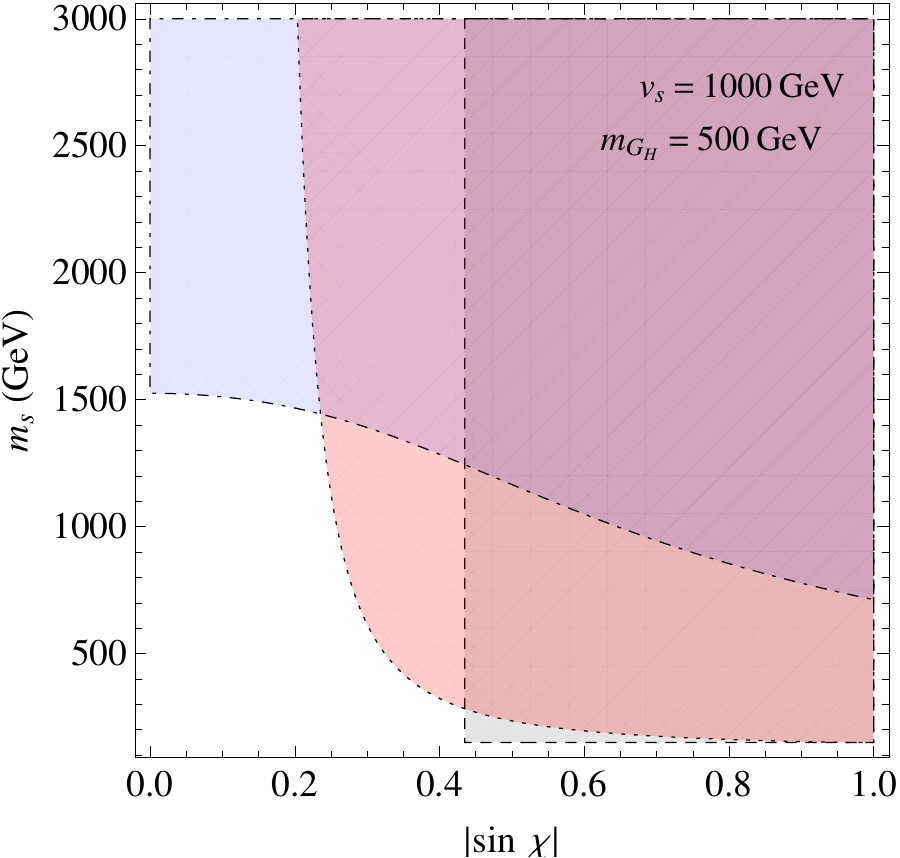}
\includegraphics[width=.329\textwidth]{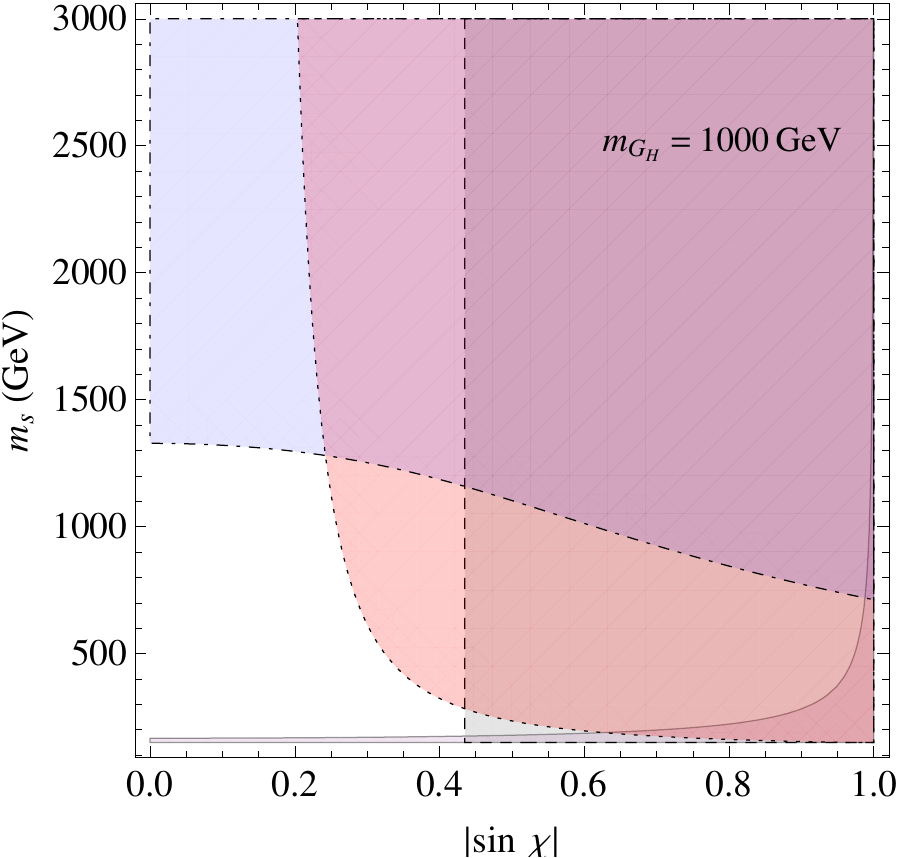}
\includegraphics[width=.329\textwidth]{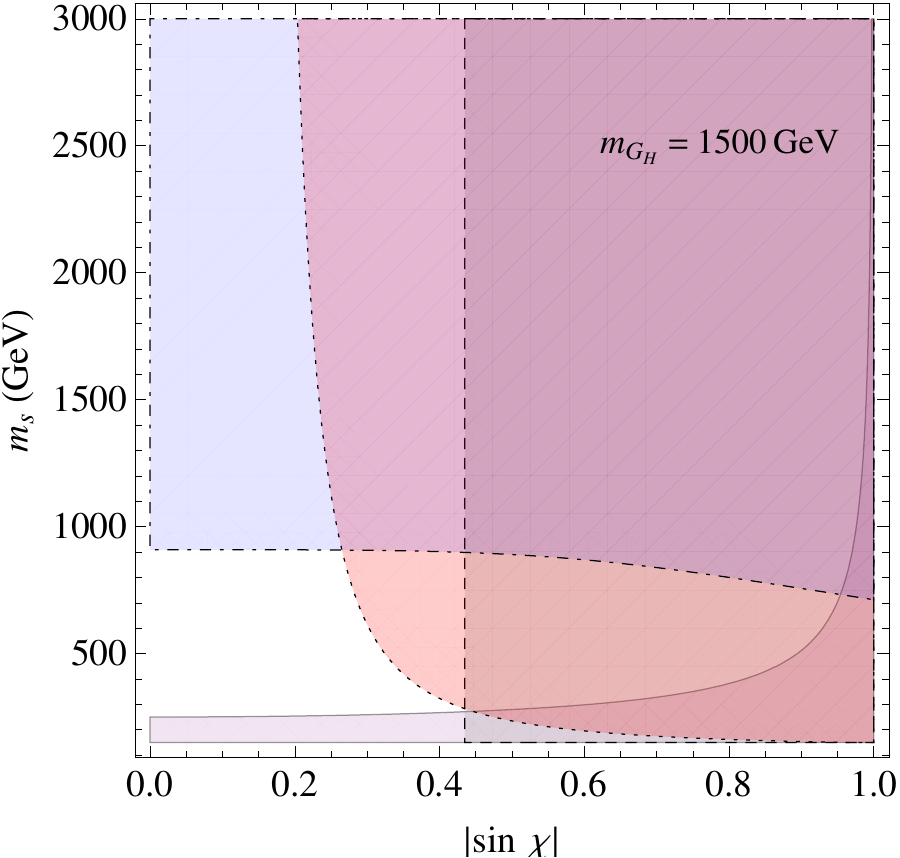}
\includegraphics[width=.329\textwidth]{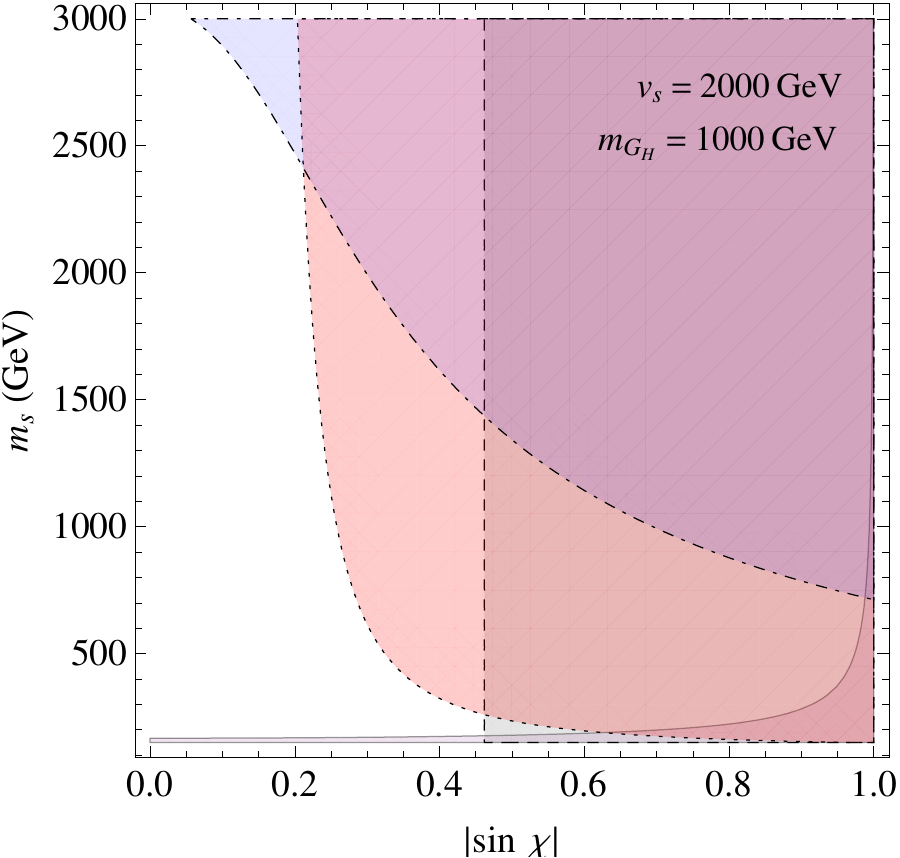}
\includegraphics[width=.329\textwidth]{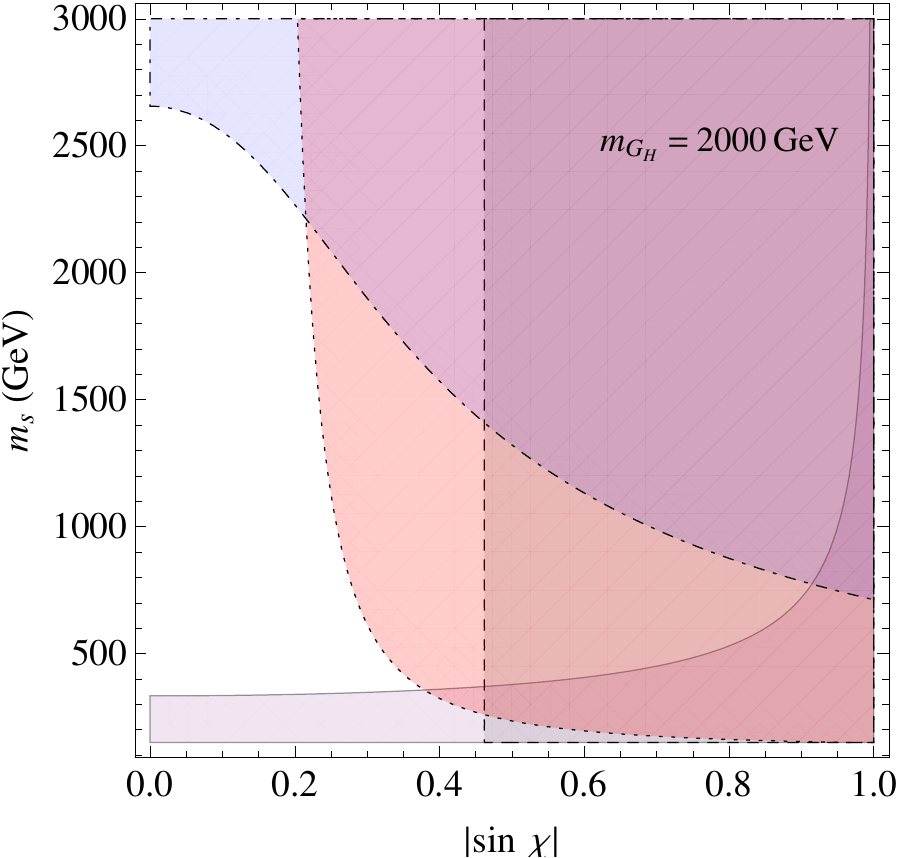}
\includegraphics[width=.329\textwidth]{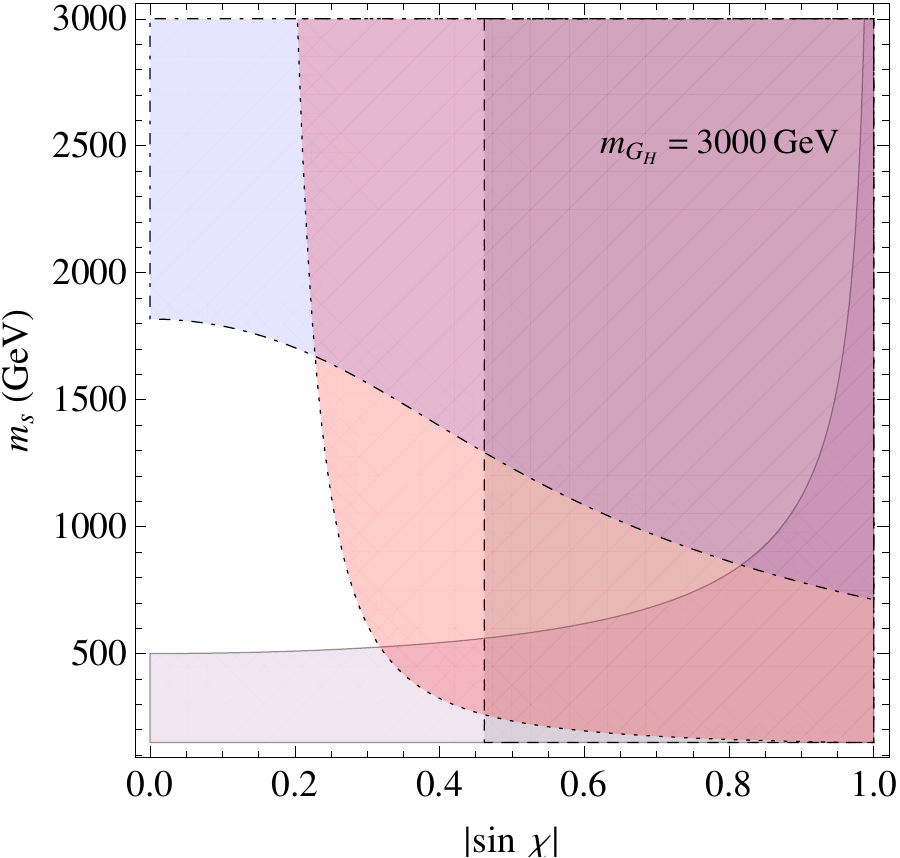}
\caption{Constraints on the $m_s- \vert\sin \chi\vert$ plane for $m_s\geq 150$~GeV. All colored regions are excluded. Theoretical bounds arise from stability (solid curves) and  unitarity (dot-dashed) analyses, whereas the experimental constraints, at 95\% C.L., are due to electroweak precision tests (dotted), and LHC direct searches (dashed, vertical). The top row shows bounds for a singlet VEV $v_s = 500$~GeV, where several values of the scalar color-octet mass, $m_{G_{H}}$, are plotted; the middle (bottom) row corresponds to $v_s = 1000\ (2000)$~GeV. In each panel, the illustrative value for the pseudo-scalar mass, $m_{\mathcal A} = \frac{1}{2} m_{G_{H}}$ is shown, since dependence on $m_{\mathcal A}$ is modest. Sensitivity to the coloron and spectator fermion masses, $M_{C}$ and $M_{Q}$ is negligible. The allowed parameter space for $v_s = 500$~GeV is tightly constrained, while a larger singlet VEV leaves more scope.}
\label{combo}
\end{figure}

In this paper, we have studied the constraints on the scalar sector of the renormalizable coloron model \cite{Hill:1993hs,Dicus:1994sw,Chivukula:1996yr,Bai:2010dj}, as introduced in section~\ref{model}, by incorporating the theoretical considerations of stability and unitarity, as well as the available experimental data on electroweak precision tests and the latest LHC results on the properties of the 125 GeV Higgs-like scalar boson. These formal and experimental analyses allowed us to significantly narrow the viable region of the parameter space, spanned by the seven input parameters of the theory \eqref{freepar}. We have displayed the results of each investigation, and the dependence of the bounds derived on $\sin\chi$, $m_s$, $m_{\mathcal A}$, and $v_s$, in the corresponding sections.

It is instructive to combine the resulting constraints, and exhibit the overall dependence of the allowed region on the input parameters. This summary analysis is displayed in Fig.~\ref{combo}, where, as before, benchmark values of the singlet VEV, $v_s = 500,\, 1000$ and $2000$ GeV, are used in the top, middle, and bottom rows.
Several values of the scalar color-octet mass, $m_{G_{H}}$, are plotted, with an illustrative pseudo-scalar mass $m_{\mathcal A} = \frac{1}{2} m_{G_{H}}$ in each plot. Because (as emphasized in section \ref{LHCsearch}) the constraints derived from LHC measurements of the properties of the light Higgs-like depend strongly on the fermion spectator content of the theory and on the sign of $\sin\chi$, the LHC bound we display in Fig.~\ref{combo} is the most conservative\footnote{That is, for a given value of $v_s$, we show as excluded only the values of $\vert\sin\chi\vert$ that are excluded for all of the possible numbers of spectator fermions treated in Fig.~\ref{fig:LHC95CLExcmA0mA100}.} bound on $\vert\sin\chi\vert$ of those shown in Fig.~\ref{fig:LHC95CLExcmA0mA100} for a given $v_s$.  The sensitivity to $m_{\mathcal A}$ is relatively mild in the entire study; the dependence on the coloron and the spectator fermion masses ($M_{C}$ and $M_{Q}$) is negligible. 

Taking the results in Fig.~\ref{combo} as a whole, we conclude that there are several interesting relationships between the constraints on the five model parameters $\{ v_s, m_s, \sin\chi, m_{\cal A}, m_{G_H} \}$:
\begin{itemize}
\item Constraints from precision electroweak tests generally restrict $\vert\sin\chi\vert$ to be less than $0.2$, with this restriction becoming (logarithmically) stronger as $m_s$ increases.
\item Stability of the potential implies that $m_s > m_{\mathcal A}/3$ and $m^2_{G_H}> 2 m^2_{\mathcal A}/3$.
\item Maintaining the unitarity of scalar boson scattering cross-sections at high energies bounds the scalar masses of
the theory for a given value of $v_s$, with $m_s/v_s \le 1.5$.
\end{itemize}
Larger values of $\vert\sin\chi\vert$ (potentiallly as large as 0.5) are also allowed, though only for relatively small values of $m_s$ and for particular, model-dependent choices for the spectator fermion content of the theory (see Fig.~\ref{fig:LHC95CLExcmA0mA100}).
Finally, in the region $\vert\sin\chi\vert \to 0$ the light scalar becomes indistinguishable from the standard model Higgs boson. Therefore, to the extent that all experimental data remain consistent with the standard model, the upper bound on $\vert\sin\chi\vert$ will become stronger over time.

We have also considered direct LHC constraints on the $s$ particle, arising from searches for a heavy Higgs boson \cite{CMSHeavyH, ATLASHeavyH}. The diboson signal rate that the $s$ would yield depends sensitively on the scalar spectrum of the theory and (especially) on the number of spectator quark generations. For $N_Q=0$ or 3, we have shown that $m_s$ is likely to be sufficiently constrained to close a substantial portion of the parameter space shown
in Fig.~\ref{fig:LHC95CLExcmA0mA100}. However, for $N_Q=1$ a cancellation between the coloron and fermionic contributions to the gluon fusion $s$-boson production cross section renders the current limits too weak to be 
definitive. 

Our analysis has also shown that the properties of the scalar sector do not constrain the masses of the colorons or the spectator fermions.  The strongest limits on these masses remain those mentioned at the start of the paper:  Tevatron and LHC data require the coloron mass, $M_{C}$, to be at least in the TeV region \cite{Simmons:1996fz,Bertram:1998wf,ATLAS:2012pu,ATLAS:2012qjz,Chatrchyan:2013qha,CMS:kxa}, while the Tevatron excludes the mass range from 50 to 125 GeV for the scalar color-octet, $m_{G_{H}}$ \cite{Aaltonen:2013hya}. 

It is fascinating to see that models with extended strong interaction sectors remain viable in light of the extensive new data from the LHC.  We look forward to seeing what discoveries might emerge in the coming years.

\section{Acknowledgments}

We thank Yang Bai for useful correspondence. We thank a referee of the manuscript for emphasizing the need to consider properties of the
$s$-boson. A.F. was supported by the Tsinghua Outstanding Postdoctoral Fellowship, and by the NSF of China (under grants 11275101, 11135003). J.R. was supported by the China Scholarship Council, and by the NSF of China (under grants 11275101, 10625522, 11135003). R.S.C. and E.H.S. were supported, in part, by the U.S.\ National Science Foundation under Grant No.\ PHY-0854889; their work was also supported in part by the National Science Foundation under Grant No. PHYS-1066293 and the hospitality of the Aspen Center for Physics.

\appendix

\section{Analyzing the Scalar Potential} \label{potanal}

In this appendix, we describe how one arrives at the form of the potential as given in Eq. \eqref{pot}.
In what follows, we modify the analysis given in 
\cite{Bai:2010dj} so as to clarify that the potential -- subject to the conditions in Eqs. \eqref{stab} and \eqref{rdel} -- has a global minimum given by Eq. \eqref{VEVs}. 

The analysis of the most general gauge-invariant quartic potential for the fields $\phi$ and $\Phi$ hinges on the dependence of the potential on the singlet field, $s_0$, defined in Eq. \eqref{Phi}. In particular,
the dependence of the potential on $s_0$ alone, $P(s_0)$, must be such that it is bounded from below and has two local minima which we parameterize as
\begin{equation}
\langle s_0\rangle = v_s,\, -v_s+\Delta~.
\end{equation}
In addition, we will assume\footnote{Note that neither of these conditions constrain
our system in any way: the first can be enforced by the transformation $s_0 \to -s_0$, and the second reflects the arbitrary choice of calling the `larger' critical point $v_s$.} that $v_s>0$ and $|v_s| > | v_s-\Delta|$, which imply 
\begin{equation}
0\le \Delta \le 2v_s \ .
\end{equation}

The most general potential satisfying these conditions
is such that the derivative of its quartic potential has the form
\begin{equation}
\frac{dP(s_{0})}{ds_0} = \lambda \,s_0 (s_0-v_s) (s_0+v_s-\Delta) \ .
\label{eq:derivative}
\end{equation}
The expression \eqref{eq:derivative} is easily integrated to yield
\begin{equation}
P(s_0) = \frac{\lambda}{4}s_0^4 - \frac{\lambda \, \Delta}{3}s_0^3 - \frac{\lambda \, v_s\pbrac{v_s-\Delta}}{2}s_0^2 \ .
\label{eq:specialpolynomial}
\end{equation}
Factoring out $\lambda \, v_s^4$, it is evident that the properties of the potential \eqref{eq:specialpolynomial}, in fact, only depend on the parameter 
\begin{equation}\label{rdelta}
r_{\Delta}\equiv \Delta/v_s \ ,
\end{equation}
once we choose the variable to be $\tilde{s}_0=s_0/v_s$. Hence, the analysis may be reduced to considering the 
behavior of the polynomial
\begin{equation}
\tilde{P}(\tilde{s}_0) = \frac{1}{4}\tilde{s}_0^4-\frac{r_{\Delta}}{3}\tilde{s}_0^3-\frac{1-r_{\Delta}}{2}\tilde{s}_0^2 \ ,
\end{equation}
as we vary $r_{\Delta}$ in the range $[0, 2]$. This polynomial has the following properties:
\begin{itemize}

\item For $r_{\Delta}=0$, the minima at $\tilde{s}_0=\pm 1$ are degenerate, which implies a massless ${\cal A}$ pseudo-scalar field (Eq. \eqref{mA});

\item For $0 < r_{\Delta} < \frac{3}{2}$, the critical point at $\tilde{s}_0=1$ is a stable, global minimum;

\item For $r_{\Delta}=\frac{3}{2}$, the minima at $\tilde{s}_0=0, 1$ are degenerate;

\item For $\frac{3}{2} < r_{\Delta} \le 2$, the global minimum is at $\tilde{s}_0=0$, corresponding to no
symmetry breaking, a physically uninteresting case.

\end{itemize}

The potential \eqref{pot} is designed in such a way that the terms involving $s_{0}$ are precisely of the form described by the polynomial \eqref{eq:specialpolynomial}, with the condition \eqref{rdel} satisfied. For this range of parameters, therefore, the global minimum of the potential $V(\phi,\Phi)$ is given by Eq. \eqref{VEVs}.

\section{Feynman Rules} \label{FR}

This appendix lists the most relevant Feynman rules of the theory. Here, the coloron is represented by a zigzag line, the gluon by the usual curly line, and the scalars by dashed lines. The spectator is depicted as a continuous double line.

\begin{figure}
\begin{center}
\includegraphics[width=.9\textwidth]{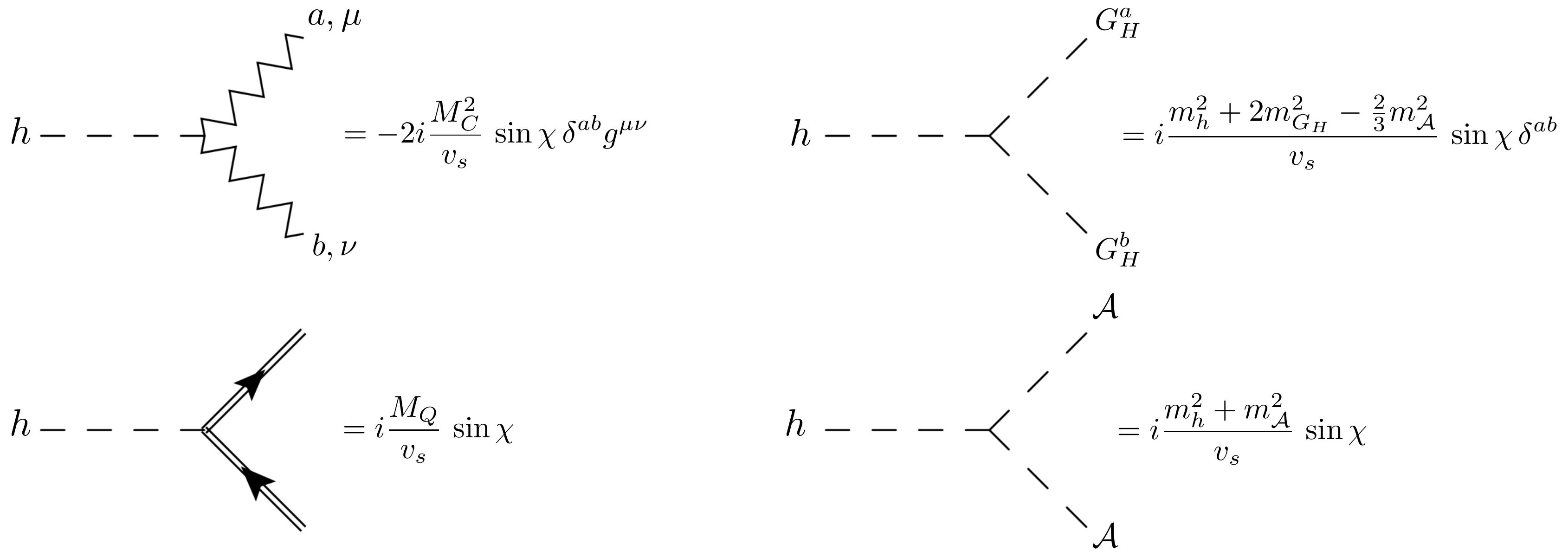}
\caption{Trilinear couplings of the $h$ boson with the coloron, the spectator fermion, and the other physical (pseudo-)scalars present in the theory.}
\label{Ch}
\end{center}
\end{figure}

\begin{figure}
\includegraphics[width=.55\textwidth]{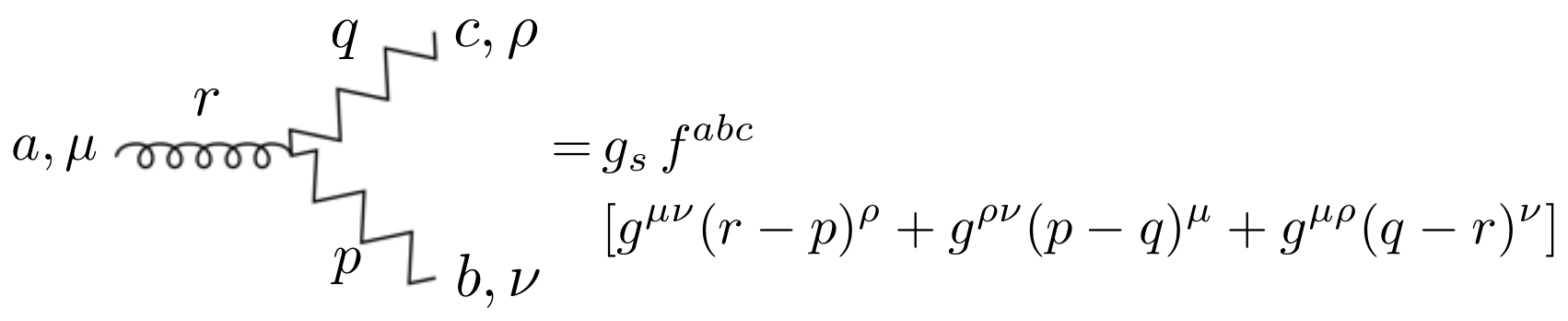} \quad
\includegraphics[width=.35\textwidth]{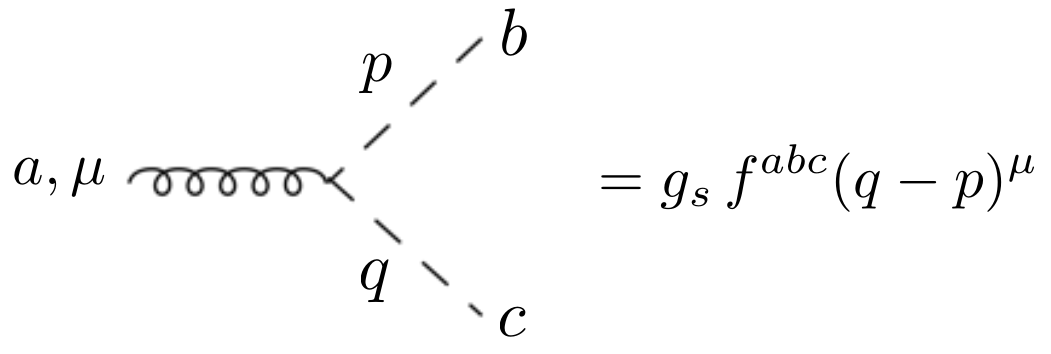}
\caption{Trilinear couplings of a gluon with a pair of vector color-octets ($C^a_\mu$), and a pair of scalar color-octets ($G_{H}^{a}$). In each diagram, all momenta flow towards the vertex.}
\label{Cg}
\end{figure}

\begin{figure}
\includegraphics[width=.45\textwidth]{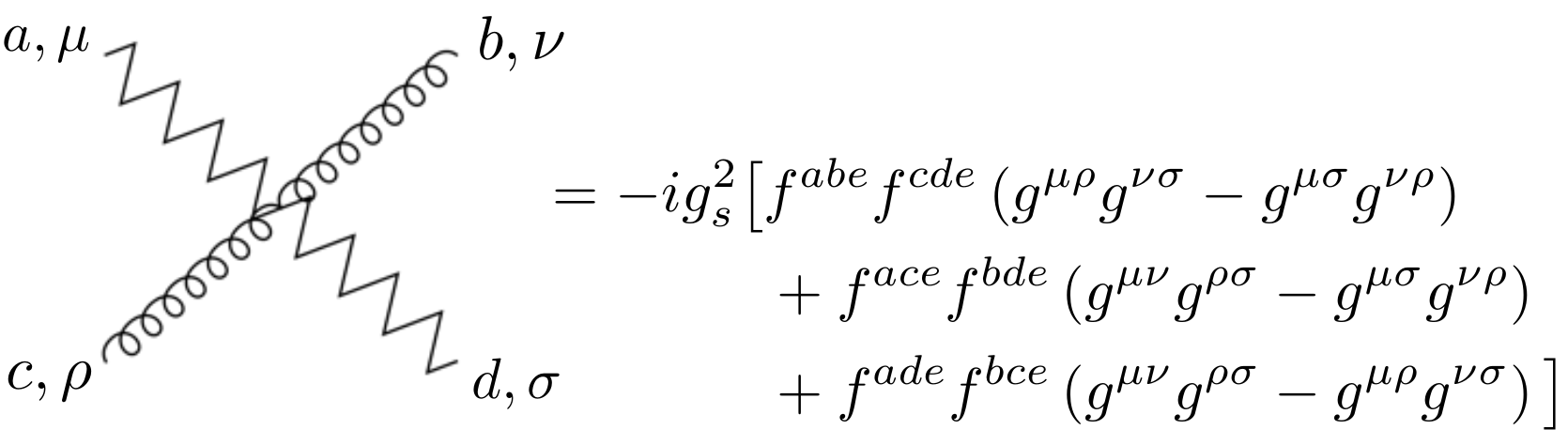} \qquad
\includegraphics[width=.45\textwidth]{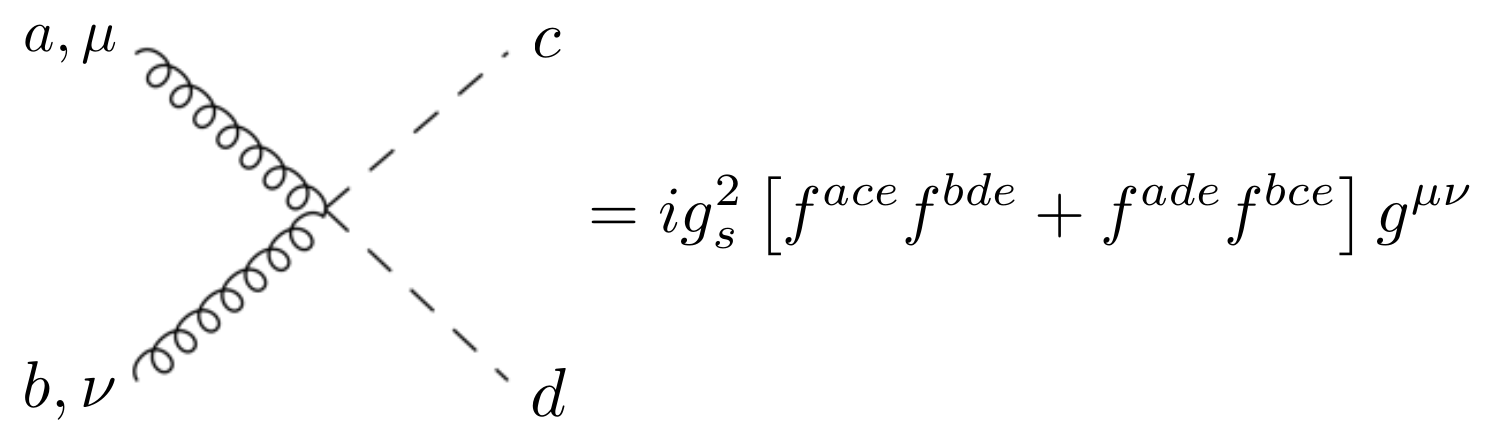}
\caption{Quartic couplings of a pair of gluons with a pair of vector color-octets ($C^a_\mu$), and a pair of scalar color-octets ($G_{H}^{a}$).}
\label{Qg}
\end{figure}

\begin{figure}
\begin{center}
\includegraphics[width=\textwidth]{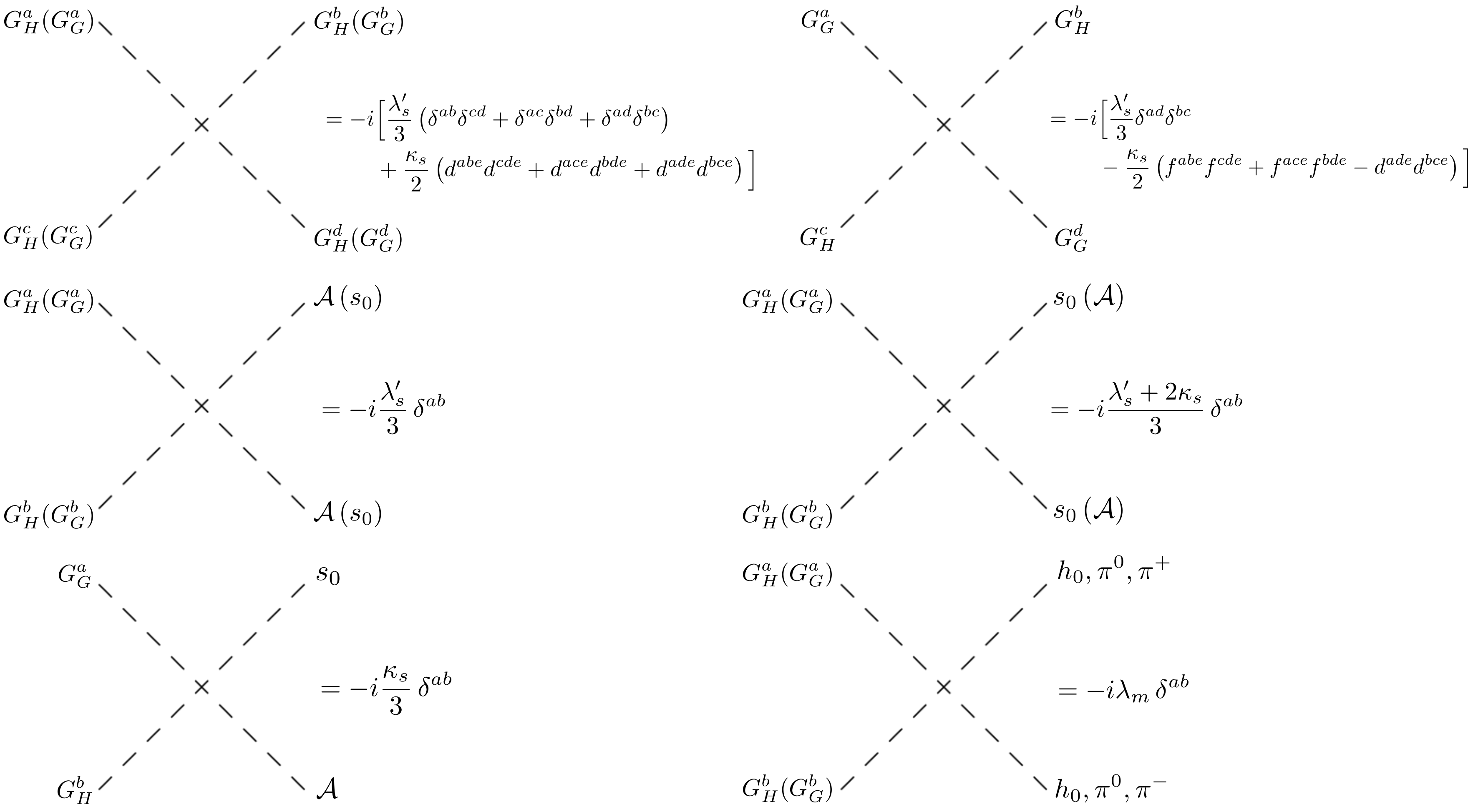}
\caption{Scalar quartic vertices, involving the colored bosons.}
\label{Qsc}
\end{center}
\end{figure}

Figs.~\ref{Ch}~and~\ref{Cg} display the relevant trilinear couplings, while the quartic interactions are shown in Figs.~\ref{Qg}~and~\ref{Qsc}. In particular, we exhibit only the non-trivial quartic couplings among the colored scalars in Fig.~\ref{Qsc}; those involving the colorless scalar states can be easily deduced from the potential \eqref{pot}.

\newpage

\includepdf[pages={1}]{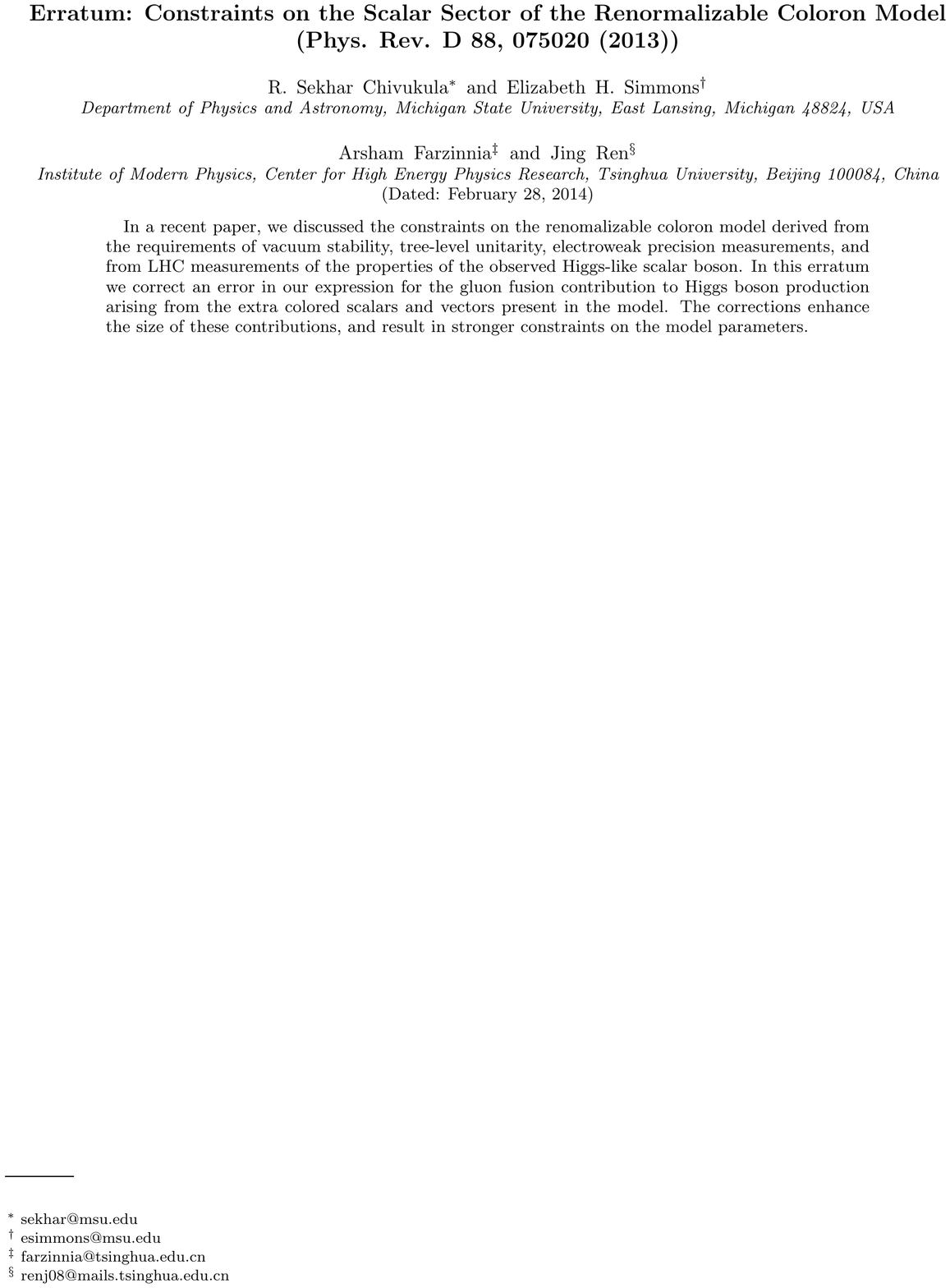}
\includepdf[pages={2}]{Erratum-final.pdf}
\includepdf[pages={3}]{Erratum-final.pdf}
\includepdf[pages={4}]{Erratum-final.pdf}
\includepdf[pages={5}]{Erratum-final.pdf}
\includepdf[pages={6}]{Erratum-final.pdf}

\end{document}